%% file: ms.tex
\documentclass[preprints,article,accept,moreauthors,pdftex]{Definitions/mdpi}
\usepackage{graphicx,color}
\usepackage{bm}
\usepackage{bbm}
\usepackage{color}
\usepackage[plain,noend]{algorithm2e}
\usepackage{graphicx}
\usepackage{amsmath,amssymb,amsthm}
\usepackage{booktabs}
\usepackage{longtable}
\usepackage{float}
\usepackage{pdflscape}

\newtheorem{alg}{Algorithm}

\input{newcommands_triple_gamma}

\input{figures_png_papers}

\firstpage{1}
\pubvolume{xx}
\issuenum{1}
\articlenumber{5}
\pubyear{2019}
\copyrightyear{2019}
\history{Received: date; Accepted: date; Published: date}

\pdfoutput=1

\Title{Triple the gamma -- A unifying shrinkage prior for variance and variable selection in sparse state space and TVP  models}

\Author{Annalisa Cadonna $^{1}$, Sylvia Frühwirth-Schnatter $^{1, *}$ and Peter Knaus $^{1}$}

\AuthorNames{Annalisa Cadonna, Sylvia Frühwirth-Schnatter and Peter Knaus}

\address{%
$^{1}$ \quad WU, Vienna University of Economics and Business}

\corres{Correspondence: sfruehwi@wu.ac.at}

\abstract{Time-varying parameter (TVP) models are very flexible in capturing gradual changes in the effect of a predictor on the outcome variable. However, in particular when the number of predictors is large, there is a known risk of overfitting and poor predictive performance, since the effect of some predictors is constant over time. We propose a prior for variance shrinkage in TVP models, called triple gamma. The
triple gamma prior encompasses a number of priors that have been suggested previously, such	as the Bayesian lasso, the double gamma prior and the Horseshoe prior. We present the desirable properties of such prior and its relationship to Bayesian Model Averaging for variance selection. The  features of the triple gamma prior are then illustrated in the context of time varying parameter vector autoregressive models, both for simulated dataset and for a series of macroeconomics variables in the Euro Area.}


\newcommand{\comment}[1]{{#1}}  

\begin{document}
\section{Introduction}
Model selection in a  high-dimensional  setting is a common challenge in statistical and econometric inference.
The introduction of Bayesian model averaging  (BMA) techniques in the statistical literature
\citep{raf-etal:bay_mod,bro-etal:bayJRSSB,cot-etal:var}
has  led to  many interesting applications, see, among  others, \citep{koo-pot:for,sal-etal:det,kle-van:baymod,fru-tue:bay} for early references in econometrics.

Predictor selection for possibly very high-dimensional regression problems though shrinkage priors    is an  attractive
alternative to BMA  which relies on discrete mixture priors, see  \citet{bha-etal:las} for an excellent review.
There is a vast and growing literature on shrinkage priors for regression problems that focuses on the following aspects.
First, how to choose sensible priors  for high-dimensional model selection problems in a Bayesian framework,
second, how to design efficient algorithms to cope with the associated computational challenges and third,
to investigate, both from a theoretical and a practical viewpoint, how  such priors  perform  in high-dimensional problems.

A striking duality exists in this very active area between Bayesian and traditional approaches.
For many shrinkage priors, the mode  of  the posterior distribution obtained in a Bayesian analysis
can be regarded as a point estimate from a regularization approach, see   \citet{fah-etal:bay} and \citet{pol-sco:loc}.
One  such example is the
popular Lasso \citep{tib:reg}  which is equivalent to a double-exponential shrinkage prior in a Bayesian context \citep{par-cas:bay}.
However, the two approaches differ  when it comes to selecting penalty parameters that impact the sparsity of the solution.
One advantage of the Bayesian framework in this context  is that the  penalty parameters are  considered to be
unknown hyperparameters which can be learned   
from the data. Such \lq\lq global-local\rq\rq\ shrinkage priors   \citep{pol-sco:shr} adjust
to the overall degree of sparsity that is required in a specific application through a global  shrinkage parameter
and separates signal from noise through local, individual shrinkage parameters.

While predictor selection though shrinkage priors in regression models is addressed in a vast literature,
the use of shrinkage priors for more general econometric models for time series analysis,
such as state space models and time-varying parameter (TVP) models is, in comparison, less well-studied.
Sparsity in the context of such models refers to the presence of a few large variances among many (nearly) zero variances in the latent state processes that drive the observed time series data.
A common goal in this setting is to recover a few dynamic states,  driven by such
a state space model, among  many (nearly) constant coefficients.
As shown by \citet{fru-wag:sto},   this variance selection problem can be cast into a variable selection problem
in the non-centered parametrization of a state space model. Once  this link has been established,  shrinkage priors that are
known to perform  well in high-dimensional regression problems can be applied to variance selection in state space models,
as demonstrated for the Lasso \citep{bel-etal:hie_tv} and the normal-gamma \citep{gri-bro:hie,bit-fru:ach}.

Despite this already existing variety, we introduce a new shrinkage prior for variance selection in
sparse state space and TVP models in the present paper, called triple gamma prior as it has a representation involving three gamma distributions.
This prior  can  be related to  various shrinkage priors that were found to be useful for   high-dimensional regression problems such as
the generalized beta  mixture prior   \citep{arm-etal:gen_bet}   and contains the popular Horseshoe prior \citep{car-etal:han,car-etal:hor} as a special case.
Furthermore,  the half-$t$  and the half  Cauchy \citep{gel:pri,pol-sco:hal},
suggested   as robust alternatives to the inverse gamma distribution  for variance parameters in hierarchical models, as well as
the Lasso and the double gamma,
are  special cases of the triple gamma. In this context, the triple gamma can also be regarded as an extension
of the  scaled beta2 distribution \citep{per-etal:sca}.

Among Bayesian shrinkage priors, usually  a clear distinction is made between two-group mixture or spike-and-slab priors  and
continuous shrinkage priors, of which the triple gamma is a special case.
An important contribution of the  present paper is to show  that the triple gamma provides a  bridge between these two approaches
and has the following property  which is favourable both  in sparse  and dense situations. One of the hyperparameters  allows high  concentration over  the region in the shrinkage profile that  is relevant for shrinking noise, while
the other  hyperparameter  allows  high  concentration over the region  that prevents overshrinking of signals. This leads to a behaviour
of the triple  gamma prior that
very much resembles Bayesian model averaging based on  discrete spike-and-slab priors, with a strong prior concentration  at the corner solutions where some of the variances are nearly close to zero.
While this is reminiscent of the Horseshoe prior, the shrinkage profile induced by the triple gamma is more flexible than that of a Horseshoe. Thanks to the estimation of the hyperparemters, it is not constrained to be symmetric around one half, enabling adaption to varying degrees of sparsity in the data. 

The triple gamma prior also scores well from a   computational perspective.
While exploring the full posterior distribution  for  spike-and-slab priors  leads to computational challenges  due to the combinatorial
complexity of the model space, Bayesian  inference   based on Markov chain Monte Carlo (MCMC) methods is straightforward for continuous
shrinkage priors, exploiting their Gaussian-scale mixture representation \citep{mak-sch:sim,bit-fru:ach}. An extension of these schemes to the triple gamma prior is  fairly straightforward.

We will study  the empirical  performance of the triple gamma for a challenging setting in econometric time series
analysis, namely for time-varying parameter vector autoregressive models with stochastic volatility (TVP-VAR-SV models).
Since the influential paper of \citet{pri:tim} (see \citet{del-pro:tim} for a corrigendum), this model
has become a benchmark for analyzing relationships between macroeconomic variables that evolve over time, see  \citet{nak:tim}, \citet{koo-kor:lar}, \citet{eis-etal:sto}, \citet{cha-eis:bay},
\citet{fel-etal:sop} and \citet{car-etal:lar}, among many others. 
Due to the high dimensionality of the time-varying parameters, even for moderately sized systems, shrinkage priors
such as the triple gamma prior are instrumental for efficient inference.

The rest of the paper is organized as follows.  In Section~\ref{sec:triplegamma}, we  define the triple gamma prior
and discuss some of its properties.
The close relationship between  the triple gamma and spike-and-slab priors applied
in a BMA context is investigated  in Section~\ref{sec_BMA}. Section~\ref{sec_mcmc} introduced an efficient MCMC scheme
and Section~\ref{applications} provides applications to TVP-VAR-SV models.
Section~\ref{conclude}  concludes the paper.

\section{The triple gamma as  a prior for  variance parameters}
\label{sec:triplegamma}

\subsection{Motivation and definition}

Let us recall the state space form of a TVP model. For $t = 1, \ldots, T$, we have that
\begin{equation} \label{eq:centeredpar}
\begin{aligned}
& \kfx_{t}  = \kfx_{t-1} + \kfw_{t}, \qquad   \kfw_t  \sim \Normult{d}{\bfz, \kfQ},\\
& y_{t}=   \Xbeta_t \kfx_{t}  +  \error_{t} , \qquad \error_{t} \sim \Normal{0,\sigma^2_t},  
\end{aligned}
\end{equation}
where $\kfQ=\Diag{\theta_1, \ldots, \theta_d}$, $y_t$ is a univariate response variable, $ \Xbeta_t = (x_{t 1},  \ldots, x_{t  d})$ is a $d$-dimensional row vector containing the regressors at time $t$, with $x_{t 1}$ corresponding to the intercept, and the initial value follows a normal distribution, $\kfx_{0} \sim \Normult{d}{\betav, \kfQ}$, with initial mean $\betav = (\beta_1, \ldots, \beta_d)^\top$.
Model (\ref{eq:centeredpar}) can be rewritten equivalently in the non-centered parametrization introduced in \citet{fru-wag:sto} as
\begin{equation}  \label{eq:noncenteredpar}
\begin{aligned}
&\btildev{t} =\btildev{t-1} + \tilde{\kfw}_{t}, \qquad \tilde{\kfw}_{t} \sim  \Normult{d}{\bfz, \identy{d}}, \\
&y_t= \Xbeta_t   \betav +  \Xbeta_t   \Diag{\sqrt{\theta_1}, \ldots, \sqrt{\theta_d}} \btildev{t}
+  \error_t, \quad  \error_t \sim \Normal{0,\sigma^2_t},
\end{aligned}
\end{equation}
with $ \btildev{0} \sim \Normult{d}{\bfz, \identy{d}} $, where $\identy{d}$ is the $d$-dimensional identity matrix.
The error variance in the observation equation is either homoscedastic ($\sigma^2_t \equiv \sigma^2$ for all $t=1,\dots,T$)  or follows a stochastic volatility (SV) specification \citep{jac-etal:bayJBES},  
where the log volatility $h_t = \log \sigma^2_t $ follows an AR(1) process. Specifically,
\begin{eqnarray} \label{svht}
h_t | h_{t-1}, \mu, \phi, \sigma_\eta^2 \sim \Normal{\mu+\phi(h_{t-1}-\mu),\sigma^2_\eta}.
\end{eqnarray}

To motivate the triple gamma prior, let us recall that, in TVP models, shrinkage priors are placed on each scale parameter $\sqrt{\theta_j}$, $j=1, \ldots,d$, in order to shrink  dynamic coefficients to static ones, hence avoiding overfitting.
One of such priors is the double gamma prior, employed recently by \cite{bit-fru:ach} for shrinkage of variances. The double gamma prior can be expressed as a scale-mixture of gamma distributions, with the following hierarchical representation:
\begin{align} \label{eq:TG}
\theta_j  | \xi_j^2\sim \Gammad{\frac{1}{2}, \frac{1}{2\xi_j^2}}, \quad
\xi_j^2| a^\xi, \kappa_B^ 2 \sim \Gammad{a^\xi, \frac{a^\xi \kappa_B^2}{2}}.
\end{align}
In the double gamma prior, each innovation variance $\theta_j$ is mixed over its own $\xi_j^2$ , each of which has an independent gamma distribution, with a common hyperparameter $\kappa_B^2$. Moreover, the parameters $\xi_j^2$ play the role of local (component specific) shrinkage parameters, while the parameter $\kappa_B^2$ is  a (common) global shrinkage parameter.

We propose an extension of the double gamma prior to a triple gamma prior, where
another layer is added to the hierarchy:
\begin{align}   \label{TRiple}
\theta_j  | \xi_j^2\sim \Gammad{\frac{1}{2}, \frac{1}{2\xi_j^2}}, \quad
\xi_j^2| a^\xi, \kappa_j^ 2 \sim \Gammad{a^\xi, \frac{a^\xi \kappa_j^2}{2}}, \quad
\kappa_j^2| c^\xi, \kappa^2_B \sim \Gammad{c^\xi, \frac{c^\xi}{\kappa^2_B}}.
\end{align}
The main difference with the double gamma prior is that the $\xi_j$ are not identically distributed, but each one depends on its component specific parameter $\kappa_j^2$.  Prior (\ref{TRiple}) contains many well-known shrinkage priors as a special case, as will be discussed in Section~\ref{sec_relations}.

To make the shrinkage behaviour of the triple gamma prior more apparent, we will work with representations that involve
the scale parameter $\sqrt{\theta_j}$, rather than  the variance $\theta_j$,  
using the fact that  $\theta_j  | \xi_j^2 \sim   \xi_j^2 \Chisqu{1}$ follows a re-scaled $\Chisqu{1}$-distribution.
If we consider both the positive and the  negative  root of $\theta_j$, then we obtain
\begin{align}   \label{TRipleII}
\sqrt{\theta_j}  | \xi_j^2 \sim    \Normal{0,\xi_j^2}, \quad
\xi_j^2| a^\xi, \kappa_j^ 2 \sim \Gammad{a^\xi, \frac{a^\xi \kappa_j^2}{2}}, \quad
\kappa_j^2| c^\xi, \kappa^2_B \sim \Gammad{c^\xi, \frac{c^\xi}{\kappa^2_B}}.
\end{align}
Hence,  prior (\ref{TRiple}) corresponds to $\sqrt{\theta_j}$ following the so-called  normal-gamma-gamma prior
consider by \citet{gri-bro:hie}  in the context of  defining hierarchical shrinkage priors for regression models.

To allow  shrinkage of dynamic coefficients toward fixed, but significant ones, we extend  \citet{bit-fru:ach}
further by assuming such a normal-gamma-gamma prior on the fixed parameter $\beta_1, \ldots, \beta_d$:
\begin{align}   \label{NGG}
\beta_j  | \tau_j^2\sim \Normal{0, \tau_j^2}, \quad
\tau_j^2| a^\tau, \lambda_j^ 2 \sim \Gammad{a^\tau, \frac{a^\tau \lambda_j^2}{2}}, \quad
\lambda_j^2| c^\tau, \lambda^2_B \sim \Gammad{c^\tau, \frac{c^\tau}{\lambda^2_B}}.
\end{align}
In   Section~\ref{sec_hyp}, we will discuss hierarchical versions of  both priors, by putting a hyperprior on the parameters
$\kappa^2_B$, $\lambda^2_B$,  $a^\xi$, $a^\tau$, $c^\xi$, and $c^\tau$.

\subsection{Properties of the triple gamma prior}

It will be shown in Theorem~\ref{theo1} that
the triple gamma prior is  a global-local shrinkage prior in the sense of  \citet{pol-sco:loc}
where the local shrinkage parameters arise from  the $\Fd{2a^\xi, 2 c^\xi}$ distribution.
This representation allows to relate the triple gamma to the well-known Horseshoe prior, see  Section~\ref{sec_relations}.
Furthermore, a closed form of  the marginal shrinkage prior $p(\sqrt{\theta_j}|\phixi,a^\xi, c^\xi)$
is given in Theorem~\ref{theo1}, which is proven in Appendix~\ref{proofs}.

\begin{Theorem}\label{theo1}
	For  the triple gamma prior defined in  (\ref{TRiple}), with $a^\xi>0$ and $c^\xi>0$,
	the following holds:
	\begin{itemize}
		\item[(a)] It has following representation as a local-global shrinkage prior:
		\begin{align}
		\label{repF} & \sqrt{\theta_j} |  \xiF _j, \kappa_B^2  \sim \Normal{0, \frac{2}{\kappa_B^2} \xiF _j },  \quad   \xiF _j | a^\xi,  c^\xi   \sim \Fd{2a^\xi, 2 c^\xi}.
		\end{align}
		\item[(b)]
		The marginal prior $p(\sqrt{\theta_j} |\phixi, a^\xi, c^\xi)$ takes the following form with $\phixi=\frac{2 c^\xi}{\kappa_B^2 a^\xi}$,
		\begin{align}  \label{theo1pd}
		p(\sqrt{\theta_j} |\phixi, a^\xi, c^\xi) =   \frac{\Gamfun{c^\xi+{\frac{1}{2}}}}{\sqrt{2\pi \phixi} \Betafun{a^\xi,c^\xi}}
		\Uhyp{c^\xi + { \frac{1}{2}}, \frac{3}{2} - a^\xi, \frac{\theta_j}{2 \phixi} },
		\end{align}
		where $\Uhyp{a,b,z}$ is the confluent hyper-geometric function of the second kind:
		\begin{align*}
		\Uhyp{a,b,z} = \frac{1}{\Gamma(a)}\int_0^\infty e^{-zt} t^{a-1}(1+t)^{b-a-1} dt.
		\end{align*}
	\end{itemize}
\end{Theorem}

\begin{figure}[t!]
	\centering
	\includegraphics[width=0.7\textwidth]{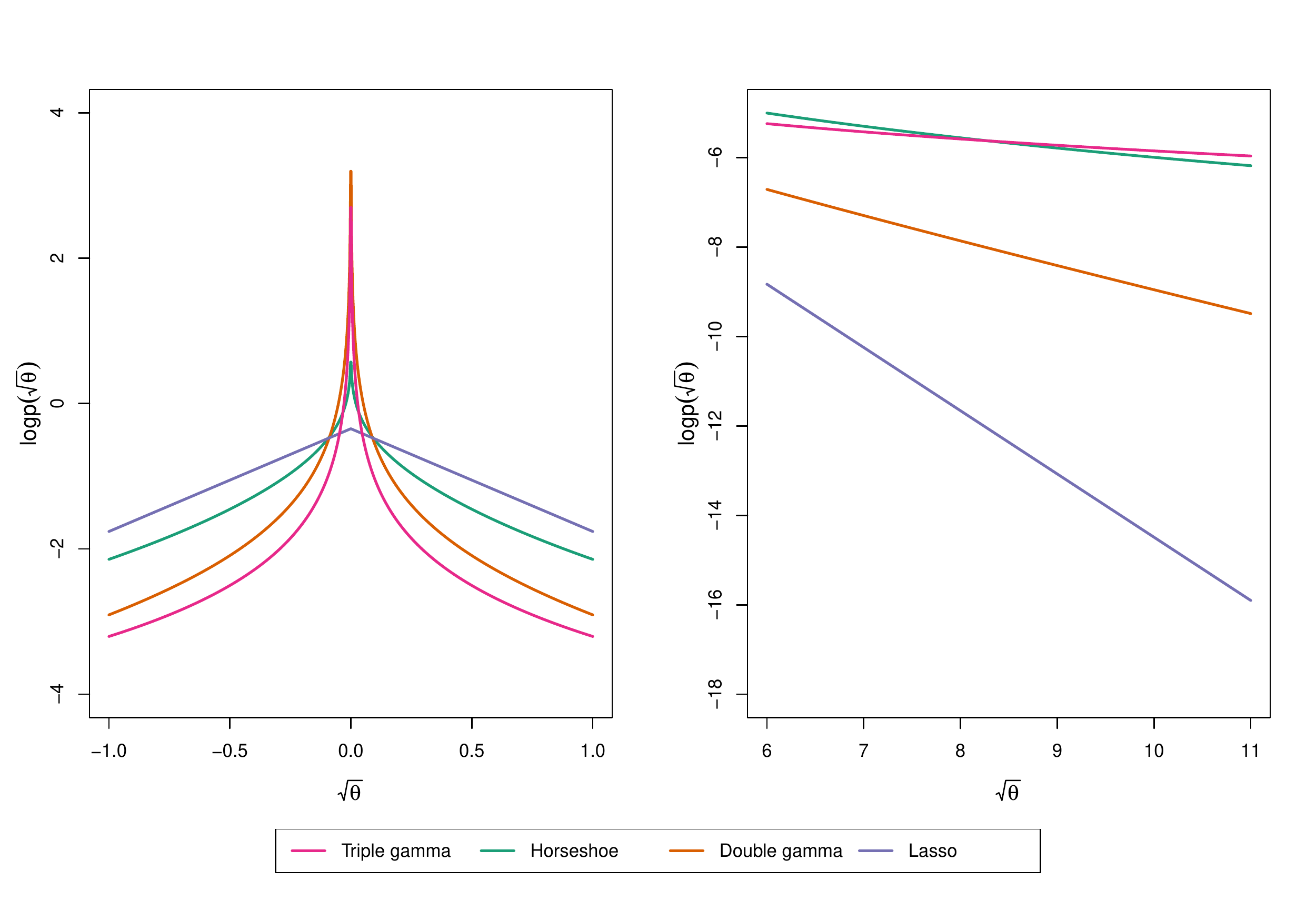}
	\caption{Marginal prior distribution of $\sqrt{\theta_j}$ under the triple gamma prior with $a^\xi=c^\xi=0.1$ with $\kappa^2_B = 2$,  in comparison to the Horseshoe prior with $\phi^\xi = 1$, the double gamma prior with $a^\xi = 0.1$ and $\kappa^2_B = 2$ and the Lasso prior with $\kappa^2_B = 2$.  (\textbf{a}) On the left, spike of the marginal prior distributions.  (\textbf{b}) On the right, tail of the marginal prior distributions. }
	\label{fig:marginals}
\end{figure}


\noindent In Figure \ref{fig:marginals} we can see the marginal prior distribution of $\sqrt{\theta_j}$ under the triple gamma prior $a^\xi=c^\xi=0.1$ and under other well-known shrinkage priors which 
are  special cases of the triple gamma, see Table~\ref{tab1}.
Using \citet[Footnote~3]{bit-fru:ach}, Theorem~\ref{theo1} also allows to give a closed
form for the prior $p(\theta_j|\phixi, a^\xi, c^\xi)= p(\sqrt{\theta_j}|\phixi, a^\xi, c^\xi)/\sqrt{\theta_j}$.

Global-local shrinkage priors are typically compared in terms of the concentration around the origin and the tail behaviour. For the triple gamma prior $p(\sqrt{\theta_j}|\phixi,a^\xi, c^\xi)$, the two shape parameters  $a^\xi$  and $c^\xi$  play a crucial  role in this respect, see
Theorem~\ref{theo4}  which is proven in Appendix~\ref{proofs}.

\begin{Theorem}\label{theo4}
	The triple gamma prior   (\ref{theo1pd}) satisfies the following:
	\begin{itemize}
		\item[(a)] For $0< a^\xi < 0.5$ and small values of $\sqrt{\theta_j}$,
		\begin{align*}
		p(\sqrt{\theta_j} |\phixi, a^\xi, c^\xi) =   \frac{\Gamfun{\frac{1}{2}- a^\xi }}{\sqrt{\pi}  (2 \phixi) ^{a^\xi} \Betafun{a^\xi,c^\xi}} \left( \frac{1}{\sqrt{\theta_j}} \right)^{1 - 2 a^\xi} + O(1).
		\end{align*}
		\item[(b)] For $a^\xi = 0.5$ and small values of $\sqrt{\theta_j}$,
		\begin{align*}
		p(\sqrt{\theta_j} |\phixi, a^\xi, c^\xi) =  \frac{1}{\sqrt{2\pi \phixi} \Betafun{a^\xi,c^\xi}}
		\left( - \log \theta_j  + \log (2 \phixi) - \psi(c^\xi+{\frac{1}{2}}) \right)
		+ O(| \theta_j \log \theta_j|),
		\end{align*}
		where $\psi(\cdot)$ is the digamma function.
		\item[(c)] For $a^\xi > 0.5$,
		\begin{align*}
		\lim_{\sqrt{\theta_j} \rightarrow 0} p(\sqrt{\theta_j} |\phixi, a^\xi, c^\xi) =
		\frac{\Gamfun{c^\xi+{\frac{1}{2}}}\Gamfun{a^\xi - \frac{1}{2}}}{\sqrt{2\pi \phixi} \Betafun{a^\xi,c^\xi}
			\Gamfun{a^\xi + c^\xi }}.
		\end{align*}
		\item[(d)] As  $\sqrt{\theta_j} \rightarrow \infty$,
		\begin{align*}
		p(\sqrt{\theta_j} |\phixi, a^\xi, c^\xi) =
		\frac{\Gamfun{c^\xi+{\frac{1}{2}}} (2 \phixi) ^{c^\xi}}{\sqrt{\pi} \Betafun{a^\xi,c^\xi}}
		\left( \frac{1}{\sqrt{\theta_j}} \right)^{ 2 c^\xi+1} \left[1  + O \left( \frac{1}{\theta_j}\right) \right].
		\end{align*}
	\end{itemize}
\end{Theorem}

\noindent  From Theorem~\ref{theo4}, Part~(a) and (b), we find that the triple gamma prior  $p(\sqrt{\theta_j}|\phixi,a^\xi, c^\xi)$  has a pole at the origin, if  $a^\xi \leq 0.5$. According to Part~(a), the pole is more pronounced, the closer $a^\xi $ gets to 0.
For $a^\xi > 0.5$, we find from Part~(c) that $p(\sqrt{\theta_j}|\phixi,a^\xi, c^\xi)$ is bounded at zero
by  a positive upper bound  which is finite,  as long as $0< c^\xi< \infty$.
Part~(d) shows that the triple gamma prior  $p(\sqrt{\theta_j}|\phixi,a^\xi, c^\xi)$ has polynomial tails, with the shape parameter $c^\xi$ controlling the tail index.
Prior moments $\Ew{(\sqrt{\theta_j})^k|\phixi,a^\xi, c^\xi}$ exist  up to $k< 2 c^\xi$. Hence,
the triple gamma prior has no finite moments for $c^\xi< 1/2$.

Finally, additional useful representations of  the triple gamma prior as a global-local shrinkage prior  are summarized in Lemma~\ref{Lem1} which is proven in Appendix~\ref{proofs}. Representations (a) 
shows that the triple gamma  is an extension of the double gamma prior where the Gaussian prior  $\sqrt{\theta_j}  | \xi_j^2 \sim    \Normaltext{0,\xi_j^2}$  is substituted by a heavier-tailed Student-$t$ prior, making the prior more robust to large values of $\sqrt{\theta_j}$.  Representation  (b) and (c) 
will be useful for MCMC inference in Section~\ref{sec_mcmc}.
Representations (c) and (d) 
show that for a triple gamma prior with  finite $a^\xi$ and $c^\xi$,
$\phixi$ 
acts as  a global shrinkage parameter, in addition to $2/\kappa_B^2$. 

\begin{Lemma}\label{Lem1}
	For $a^\xi>0$ and $c^\xi>0$, the triple gamma prior (\ref{TRiple}) has the following alternative representations:
	\begin{align}
	\label{repST} & \text{(a)} \quad \sqrt{\theta_j} |  \xitilde_j ^2,c^\xi, \kappa_B^2 \sim \Student{2c^\xi}{0, \frac{2}{\kappa_B^2} \xitilde_j^2},
	\quad  \xitilde_j^2| a^\xi   \sim \Gammad{a^\xi,  a^\xi},\\
	\label{repSTphi} & \text{(b)} \quad  \sqrt{\theta_j} |  \xiphi_j ^2,c^\xi, \kappa_B^2  \sim \Student{2c^\xi}{0,  \frac{2}{a^\xi \kappa_B^2}  \xiphi_j^2},
	\quad  \xiphi_j^2| a^\xi   \sim \Gammad{a^\xi, 1}.
	\end{align}
	Additional representations for $0<a^\xi<\infty$ and $0<c^\xi<\infty$ based on  $\phixi=\frac{2 c^\xi}{\kappa_B^2 a^\xi}$
	are
	\begin{align}
	\label{TRipleSqphi}  &  \text{(c)} \quad  \sqrt{\theta_j} |  \xiphi_j ^2 , \kappaphi_j^2, \phixi \sim \Normal{0, \phixi
		\xiphi_j^2/\kappaphi_j^2},  
	\quad  \xiphi_j^2| a^\xi   \sim \Gammad{a^\xi,  1}, \quad \kappaphi_j^2| c^\xi  \sim \Gammad{c^\xi,  1} ,\\
	\label{repBP} & \text{(d)} \quad  \sqrt{\theta_j} | \xiFtilde _j, \phixi   \sim \Normal{0, \phixi \, \xiFtilde _j },  \quad   \xiFtilde _j | a^\xi,  c^\xi   \sim \Betapr{a^\xi,c^\xi},
	\end{align}
	where  $\Betapr{a^\xi,c^\xi}$ is the beta-prime distribution.\footnote{Note that the $ X \sim \Betapr{a ,b}$-distribution has pdf
		$$
		p(x )= \frac{1}{\Betafun{a,b}}  \frac{x ^{a-1}}{(1+x)^{a+b}}.$$ Furthermore, $Y= X/(1+X)$ follows the
		$ \Betadis{a ,b}$-distribution.}
\end{Lemma}

\subsection{Relation of the triple gamma to other shrinkage priors} \label{sec_relations}

The triple gamma prior can be related to the very active research on shrinkage priors in a Bayesian framework
in various ways. 
On the one hand, 
popular   priors for variance parameters   introduced as robust alternatives to  the inverse gamma prior
are  special cases of the triple gamma, see Table~\ref{tab1}.
For instance, in  (\ref{repF}),  $\xiF_j$  converges a.s. to 1, as  $a^\xi \to \infty$ and $c^\xi \to \infty$,
and  the triple gamma reduces to a normal distribution for $\sqrt \theta _j$,  applied
for  univariate  TVP models  \citep{fru:eff}  and unobserved component state space model
\citep{fru-wag:sto}. For $c^\xi \to \infty$,  $\Fd{2a^\xi, 2 c^\xi} $  converges  to the $\Gammad{a^\xi,a^\xi}$  distribution  and   the triple gamma reduces    to  the Bayesian Lasso for $a^\xi=1$
\citep{bel-etal:hie_tv}   and  otherwise
to the double gamma   \citep{bit-fru:ach} applied  in sparse TVP models.

\citet{gel:pri} introduced the half-$t$  and the half-Cauchy prior for  variance parameters in hierarchical models,
by assuming that $\sqrt \theta _j$ follows a \lq\lq folded\rq\rq\ $t$-distribution, i.e. a $t$-distribution truncated to $[0, \infty)$,
see also  \citet{pol-sco:hal}.
In (\ref{repST}),   $\tilde{\xi}_j ^2$ converges a.s. to 1 as  $a^\xi \to \infty$
and  the triple gamma reduces to a  $\Studentnu{2 c^\xi}$- distribution and  to the Cauchy distribution  for $ c^\xi =1/2$,
however without being \lq\lq folded\rq\rq , since we allow $\sqrt \theta _j$ to take on negative values.
A half-$t_{\nu}$ with $\nu= 2 c^\xi$ and a triple gamma with $a^\xi=\infty$ obviously imply the same prior for $\theta _j$,
so  does the negative half matter?   It matters, whenever inference is performed in a parametrization involving $\sqrt \theta _j$ such as the non-centered parametrization (\ref{eq:noncenteredpar}). Restricting the prior to the positive half will lead to automatic truncation of
the full conditional posterior $p(\sqrt \theta_j| \btildev{0}, \ldots, \btildev{T}, \ym, \cdot)$, e.g. during MCMC sampling (see Section~\ref{sec_mcmc}). If the positive and the negative mode of the marginal posterior  $p(\sqrt \theta_j| \ym)$ are well-separated, then this will not matter. However, if
the true value of $\theta _j$ is close to or equal to zero, then $p(\sqrt \theta_j| \ym)$ is concentrated at zero and
truncation at 0 will introduce a bias, because the negative half is missing.

%

\begin{table}[t!]
	\centering
	\begin{tabular}{llcccc}
		\hline
		Prior for  $ \sqrt{\theta}_j  $   &  & $a^\xi$  & $c^\xi$  & $\kappa_B^2$  & $\phi^\xi$ \\[.5mm]
		\hline
		$\Normal{0,  \xiF_j },   \xiF_j \sim \GG{a^\xi,c^\xi,\phi^\xi}$ & normal-gamma-gamma  &   $ a^\xi$   & $ c^\xi$   & $\frac{2c^\xi}{\phi^\xi a ^\xi}$  & $\phi^\xi$ \\[.5mm]
		$ \Normal{0, \frac{1}{\kappa_j}-1}, \kappa_j   \sim \TPB{a^\xi, c^\xi, \phi^\xi}$
		& generalized beta mixture &   $ a^\xi$   & $ c^\xi$   & $\frac{2c^\xi}{\phi^\xi a ^\xi}$  & $\phi^\xi$ \\[.5mm]
		$\Normal{0,  \xiF_j },   \xiF_j \sim \SBeta{a^\xi, c^\xi, \phi^\xi}$ & hierarchical scaled beta2   &   $ a^\xi$   & $ c^\xi$   & $\frac{2c^\xi}{\phi^\xi a ^\xi}$   & $\phi^\xi$  \\[.5mm]
		\hline
		$ \DE{0, \sqrt{2}\, \psi _j},   \xiF  _j  \sim \Gammad{c^\xi, \frac{1}{\lambda^2}}$ & normal-exponential-gamma
		&  $1 $  & $ c^\xi$  & $2 \lambda^2 c^\xi$  & $\frac{1}{\lambda^2}$   \\[.5mm]
		$  \Normal{0, \tau^2 \xiF_j},   \xiFsqr_j \sim \Studentnu{1}$&  Horseshoe
		&   $ \frac{1}{2}$  &  $ \frac{1}{2}$  &  $\frac{2}{\tau^2}$   &  $\tau^2$ \\[.5mm]
		$ \Normal{0, \frac{1}{\kappa_j}-1}, \kappa_j   \sim \Betadis{1/2,1} $
		& Strawderman-Berger &   $ \frac{1}{2}$   & 1  & 4 & 1 \\[.5mm]
		$  \Normal{0, \tau^2 \xitilde_j },   \xitilde_j \sim \Gammad{a^\xi,  a^\xi} $& double gamma
		&  $ a^\xi$  & $ \infty$  &  $\frac{2}{\tau^2}$   & -\\[.5mm]
		$  \Normal{0, \tau^2 \xitilde_j },  \xitilde_j \sim \Exp{1} $&  Lasso   &  $ 1$  & $ \infty$  &  $\frac{2}{\tau^2}$ & - \\[.5mm]
		$  \Student{\nu}{0,\tau^2}$ &   half-$t$    & $\infty$  & $ \frac{\nu}{2}$  & $\frac{2}{\tau^2}$ & - \\[.8mm]
		$ \Student{1}{0,\tau^2}$ &   half-Cauchy       &  $\infty$  & $ \frac{1}{2}$  &  $\frac{2}{\tau^2}$  & - \\[.5mm]
		$  \Normal{0, B_0} $      & normal     & $\infty$  & $ \infty$  & $\frac{2}{B_0}$  & -\\[.8mm]
		\hline
	\end{tabular}
	\caption{Priors on  $\sqrt{\theta}_j$ 
		which are  equivalent to  (top) or  special cases of (bottom) the triple gamma prior.}\label{tab1}
\end{table}

On the other hand, the triple gamma prior   is related to  popular shrinkage priors  in regression models.
It extends the generalized beta  mixture prior
introduced  by \citet{arm-etal:gen_bet}  for variable selection in regression models,
\begin{align*} 
\beta_j  |  \xi_j^2 \sim \Normal{0,   \xi_j^2}, \quad
\xi_j^2   \sim \Gammad{a^\xi, \lambda_j}, \quad
\lambda_j  \sim \Gammad{c^\xi, \phi^\xi},  
\end{align*}
to variance selection in state space and TVP models.
This is evident from  rewriting (\ref{TRiple}) as
$ \xi_j^2   \sim \Gammad{a^\xi, \lambda_j}, 
\lambda_j  \sim \Gammad{c^\xi, \phixi}$.
%
We  exploit this relationship  in Section~\ref{sec_shriprofile} to investigate the  shrinkage profile of a triple gamma prior.
Using   \citet[Definition~2]{arm-etal:gen_bet}, the triple gamma prior can  be written as
\begin{align} \label{tpbn2}
\sqrt \theta_j   | \kappashr_j \sim \Normal{0, 1/\kappashr_j-1}, \quad
\kappashr_j| a^\xi, c^\xi, \phixi \sim \TPB{a^\xi, c^\xi, \phixi},
\end{align}
where   $\TPB{a^\xi, c^\xi, \phixi}$ is the three-parameter beta (TPB) distribution with density:
\begin{eqnarray} \label{densFrho}
p(\kappashr_j )= \frac{1 }{\Betafun{a^\xi,c^\xi}}(\phixi) ^{c^\xi}  \kappashr_j   ^{c^\xi -1}
(1- \kappashr_j )  ^{a^\xi -1}   \left( 1+ (\phixi-1)\kappashr_j \right) ^{-(a^\xi+c^\xi)}.
\end{eqnarray}
 From  (\ref{tpbn2}) and (\ref{densFrho}),  it becomes  evident that the Strawderman-Berger prior  $\sqrt \theta_j \sim \Normal{0, 1/\kappashr_j-1}$,
$\kappashr_j \sim  \Betadis{1/2,1} $ \citep{str:pro,ber:rob} is that special case of the triple gamma prior where $\phixi=1$, $a^\xi=1/2$, and  $c^\xi=1$.

The special case  of  a triple gamma, where $a^\xi=c^\xi =1/2$, corresponds to
a Horseshoe prior \citep{car-etal:han,car-etal:hor} on $\sqrt \theta_j $ with global shrinkage parameter $\tau^2=2/\kappa_B^2$,
since  $\xiF _j    \sim \Fd{1,1}$ implies that $ \xiFsqr _j  \sim \Studentnu{1}$. 
The Horseshoe prior has been introduced  for variable selection in regression models and has been shown
to have excellent theoretical properties in this context  for the \lq\lq nearly black\rq\rq\ case \citep{van-etal:hor}.
The triple gamma is  a generalization of the Horseshoe prior, with a similar shrinkage profile,
however with much more mass close to  the corner solutions.
Most importantly, as will be discussed in Section~\ref{sec_shriprofile}, this leads to a BMA-type  behaviour of the triple gamma prior for small values of $a^\xi$ and $c^\xi$.


The vast literature on shrinkage priors contains many more related priors.
Rescaling  $ \xi_j^2=  2/(\kappa_B^2) \xiF _j$  in  (\ref{repF}), for instance, yields
a representation involving a scaled beta2 distribution,\footnote{The  pdf of  a  $\SBeta{a, c, \phi}$-distribution reads:
	\begin{eqnarray*} 
		p( x )=  \frac{1}{\phi ^{a } \Betafun{a , c}} \!
		x  ^{a -1}\!  ( 1+ x/\phi) ^{-(a +c)},
	\end{eqnarray*}}
	\begin{eqnarray} \label{densFpsi}
	\sqrt \theta_j |\xi_j^2  \sim \Normal{0,\xi_j^2}, \qquad  \xi_j^2| a^\xi, c^\xi, \phixi  \sim \SBeta{a^\xi, c^\xi, \phixi}, 
	\end{eqnarray}
	as  is easily derived from (\ref{densF}).
	The scaled beta2 was introduced by \citet{per-etal:sca} in hierarchical models as a robust prior  for scale parameters,  $\sqrt \theta_j $,
	and variance parameters, $\theta_j$,  alike.
	Based on   (\ref{densFpsi}), the triple gamma can be seen as a hierarchical extension of  this  prior which  puts  a
	scaled beta2 distribution  on the scaling parameter  $\xi_j^2$ of a  Gaussian prior for $\sqrt \theta_j $, see Table~\ref{tab1}.
	\citet{gri-bro:hie} termed  prior (\ref{densFpsi})   gamma-gamma distribution, denoted by $\GG{a^\xi,c^\xi,\phi}$.
	
	For  $a^\xi=1$,  the triple gamma  reduces to the normal-exponential-gamma
	which   has a representation as a scale-mixture of   double exponential  $\DE{0, \sqrt{2} \psi _j}$-distributions, 
	see  Table~\ref{tab1}.
	It   has been considered for variable selection in regression models \citep{gri-bro:bay}  and
	locally adaptive B-spline models \citep{sch-kne:loc}.
	The R2-D2 prior suggested by \citet{zha-etal:hig} for high-dimensional regression models is another special case of the triple gamma.
	It reads
	\begin{align*} 
	\beta_j   \sim \Normal{0, \sigma^2 \phi _j \omega},  \quad  (\phi _1, \ldots,\phi _\betad ) \sim \Dir{ a^\tau, \ldots,   a^\tau},  \quad
	\omega \sim  \Gammad{a, \tau}, \quad \tau \sim \Gammad{b, 1},
	\end{align*}
	where   $a=\betad  a^\tau$ and $\sigma^2$ is the residual error variance of  the regression model.
	As shown by \citet{zha-etal:hig},   this   implies following  prior for  the coefficient of determination: $ R^2 \sim \Betadis{a,b}$
	which motivates holding $a$ fixed, while $a^\tau$ decrease as $d$ increases.
	Using  that $ \phi _j \omega  \sim  \Gammad{a^\tau, \tau}$,
	we can show that the  R2-D2 prior is equivalent  to following  hierarchical normal gamma prior
	applied in \citet{bit-fru:ach} for  TVP models:
	\begin{align*}  
	\beta_j | \tau^2_j \sim \Normal{0,  \tau^2_j},  \quad  \tau^2_j \sim  \Gammad{a^\tau, a^\tau \lambda_B^2/2},
	\quad  \lambda_B^2  \sim \Gammad{b, 2 \sigma^2 /a^\tau}.
	\end{align*}
	The  popular Dirichlet-Laplace prior,  $\sqrt \theta_j | \psi _j \sim  \DE{0, \psi _j}$,
	however,  is not related to the triple gamma  as   the {\em  prior  scale}   $\psi   _j$
	rather than   $\xiF _j$ follows a gamma distribution, see  again Table~\ref{tab1}.
	
	\subsection{Using the triple gamma for variance  selection in TVP models}  \label{sec_hyp}
	

	A challenging question is how to choose the parameters
	$a^\xi$,  $ c^\xi$  and $\kappa_B^2$ or
	$\phixi$  of  the triple gamma prior in the context of variance selection  for TVP models.
	In addition, in a TVP context, the shrinkage parameters
	$a^\tau$, $ c^\tau$  and $\lambda_B^2$ or
	$\phitau=2 c^\tau /( a^\tau \lambda_B^2) $ for the prior (\ref{NGG}) of the initial values have to be selected.
	
	In high-dimensional settings it is appealing to have a prior that addresses two major issues:
	first, high concentration around the origin to favor strong shrinkage of small
	variances toward zero; second, heavy tails to introduce  robustness  to  large  variances and to avoid over-shrinkage.
	For the triple gamma prior, both issues are addressed
	through the choice of $a^\xi$ and   $ c^\xi$, see Theorem~\ref{theo4}. First of all, we need values $ 0 < a^\xi \leq 0.5$  to induce a pole at 0.  Second,
	values of $0 < c^\xi < 0.5 $ will lead to very heavy tails.
	For very small values of $a^\xi$ and   $ c^\xi$, the triple Gamma is a proper prior that behaves nearly as the improper normal-Jeffrey's prior \citep{fig:ada}, where  $p(\sqrt \theta_j ) \propto 1/\sqrt \theta_j $ and
	$p(\kappashr_j ) \propto  \kappashr_j^{-1} (1- \kappashr_j)^{-1}$.

	Ideally, we  would place a hyper prior distribution on  all shrinkage parameters
	which would allow us to learn  the global  and  the local degree of  sparsity,  both for the variances and the initial values.
	Such a hierarchical triple gamma prior  introduces dependence among the local shrinkage parameters $\xi^2_1, \ldots, \xi^2_d$  in (\ref{TRiple})
	and, consequently, among $\theta_1, \ldots, \theta_d$ in the joint (marginal) prior  $p(\theta_1, \ldots, \theta_d)$.
	Introducing such dependence is desirable in that it allows to learn the degree of variance sparsity in TVP models,
	meaning that how much a variance is shrunken toward zero depends on how close the other  variances are to zero.
	However, first na\"{i}ve approaches with
	rather uninformative, independent priors on $\kappa_B^2$,  $a^\xi$, $ c^\xi$
	and $\lambda_B^2$, $a^\tau$, $ c^\tau$
	were not met with much success and we found it necessary to carefully design appropriate hyper priors.
	
	Hierarchical versions of  the Bayesian Lasso  \citep{bel-etal:hie_tv} and the double gamma prior \citep{bit-fru:ach} in TVP models
	are
	based on the gamma prior $ \kappa_B^2  \sim \Gammad{d_1, d_2}$. Interestingly, this choice  can be seen as a heavy-tailed extension  of both priors,
	where each marginal density  $p(\sqrt \theta_j  |d_1, d_2)$  follows a triple gamma prior with the same parameter  $a^\xi$ (being equal to one for the Bayesian Lasso) and  tail index $c^\xi =d_1$.
	In light of this relationship, it is not surprising that  very small values of $d_1$ were applied in these papers  to ensure heavy tails
	of $p(\sqrt \theta_j  |d_1, d_2)$.
	Since a triple gamma prior has already heavy tails, we choose a different  hyperprior  in  the present paper.
	
	For the case $a^\xi=c^\xi =1/2$,  the  global shrinkage parameter $\tau$  of the Horseshoe prior  typically follows a  Cauchy prior, $\tau \sim \Studentnu{1}$    \citep{car-etal:han,bha-etal:hor_reg}, see  also \citet[Section~5]{bha-etal:las}.
	The relationship $\phixi = 2/\kappa_B^2= \tau^2 $ between the  various global shrinkage parameters  (see Table~\ref{tab1})
	implies  in this case $\phixi \sim   \Fd{1,1}$  or,  equivalently, $\kappa^2_B/2  \sim   \Fd{1,1}$.
	
	For a  triple gamma prior  with arbitrary $a^\xi$ and $c^\xi$, this is a special case of the following prior:
	\begin{align}  \label{hypfinal}
	\left.  \frac{\kappa_B^2}{2}  \right|  a^\xi , c^\xi   \sim \Fd{2a^\xi ,2 c^\xi}  ,
	\end{align}
	which will be motivated in Section~\ref{sec_BMA}.  Under this prior,  the  triple gamma prior  exhibits a BMA-like behavior  with a uniform prior on an  appropriately defined model size (see Theorem~\ref{theo2}).
	Prior (\ref{hypfinal}) is equivalent with following representations:
	\begin{align}
	\label{hypkappa} & \kappa_B^2   |  a^\xi   \sim \Gammad{a^\xi, d_2}, \qquad  d_2  | a^\xi, c^\xi  \sim \Gammad{c^\xi, \dfrac{2c^\xi}{a^\xi} },   \\
	&  \phixi   |  a^\xi , c^\xi   \sim \Betapr{c^\xi,a^\xi }.  \nonumber  
	\end{align}
	

	\noindent Concerning $a^\xi $ and  $ c^\xi$, we choose the following priors:
	\begin{align}   \label{prioraxi}
	2 a^\xi \sim \mathcal B (\alpha_{a^\xi},  \beta_{a^\xi}) ,
	\qquad 2 c^\xi \sim \mathcal B (\alpha_{c^\xi},  \beta_{c^\xi}).
	\end{align}
	Hence, we are restricting the support of $a^\xi $ and  $ c^\xi$ to $(0, 0.5)$, following the insights brought to us by Theorem \ref{theo4}.

	
	We follow a similar strategy for the parameters  $a^\tau$, $ c^\tau$  and $\lambda_B^2$ ($\phitau$) of the prior (\ref{NGG}):
	\begin{align}     \label{priortau}
	\left. \frac{\lambda_B^2}{2} \right| a^\tau , c^\tau   \sim \Fd{2a^\tau ,2 c^\tau}   ,  	
\quad \comment{2 a^\tau \sim \mathcal B (\alpha_{a^\tau},  \beta_{a^\tau}) ,
	\quad 2 c^\tau  \sim \mathcal B (\alpha_{c^\tau},  \beta_{c^\tau}),}
	\end{align}
	which is equivalent with
	$\lambda_B^2   |  a^\tau   \sim \Gammad{a^\tau, e_2}$, $ e_2  | c^\tau  \sim \Gammad{c^\tau, 2c^\tau/a^\tau}$,
	and $\phitau   |  a^\tau , c^\tau  \sim \Betapr{c^\tau,a^\tau }$.

	An interesting special case  is  the  \lq\lq symmetric\rq\rq\  triple gamma, where $a^\xi = c^\xi$.
	Despite this constraint, the favourable shrinkage behaviour is preserved  and
	decreasing $a^\xi=c^\xi$  toward zero  simultaneously  leads to a high concentration around the origin and a heavy-tailed behaviour.
	For a  symmetric triple gamma prior,  $\phixi=2/\kappa_B^2$ is independent of $a^\xi$ and $ c^\xi$
	and the two global shrinkage parameters are related through $\phixi=2/\kappa_B^2$. This induces
	shrinkage profiles that are symmetric around 1/2, see Section~\ref{sec_shriprofile}.
	Interestingly,  a symmetric  triple gamma   resolves the question  whether to choose  a gamma or an inverse gamma prior for a  variance  parameter $ \xiF _j$. It  implies the same  symmetric beta-prime distribution
	on  the variance, $ \xiF _j \sim \Fd{2a^\xi, 2 a^\xi}=\Betapr{a^\xi,a^\xi}$,
	and  the  information, $( \xiF _j)^{-1} \sim \Betapr{a^\xi,a^\xi} $,
	and  can be represented as a gamma prior with the  scale arising from an inverse gamma prior
	or, equivalently, as an inverse gamma prior with the scale  arising  from a gamma prior:
	\begin{eqnarray*}
		&& \xiF _j   = \xiphi_j^2  \times \frac{1}{ \kappaphi_j^2},   \quad     (\xiF _j)^{-1}   =\kappaphi_j^2
		\times \frac{1}{\xiphi_j^2 },   \qquad \xiphi_j^2 \sim \Gammad{a^\xi, 1},  \,\, \kappaphi_j^2 \sim  \Gammad{a^\xi,  1}.
	\end{eqnarray*}

	
	\section{Shrinkage profiles and BMA-like behavior}    \label{sec_pro_BMA}
	
	\subsection{Shrinkage profiles}   \label{sec_shriprofile}

	In the sparse normal-means problem where
	$ \ym|\betav \sim \Normult{\betad}{\betav, \sigma^2 \identy{\betad}}$
	and  $\sigma^2=1$,  the parameter  $\kappashr_j=1/(1+ \xiF_j)$  appearing in (\ref{tpbn2})  is known as shrinkage factor  and plays a fundamental role
	for comparing
	different shrinkage priors, as  $\kappashr_j$ determines shrinkage toward 0. 
	
\begin{figure}[t!]
			\centering
			\includegraphics[width=0.5\textwidth]{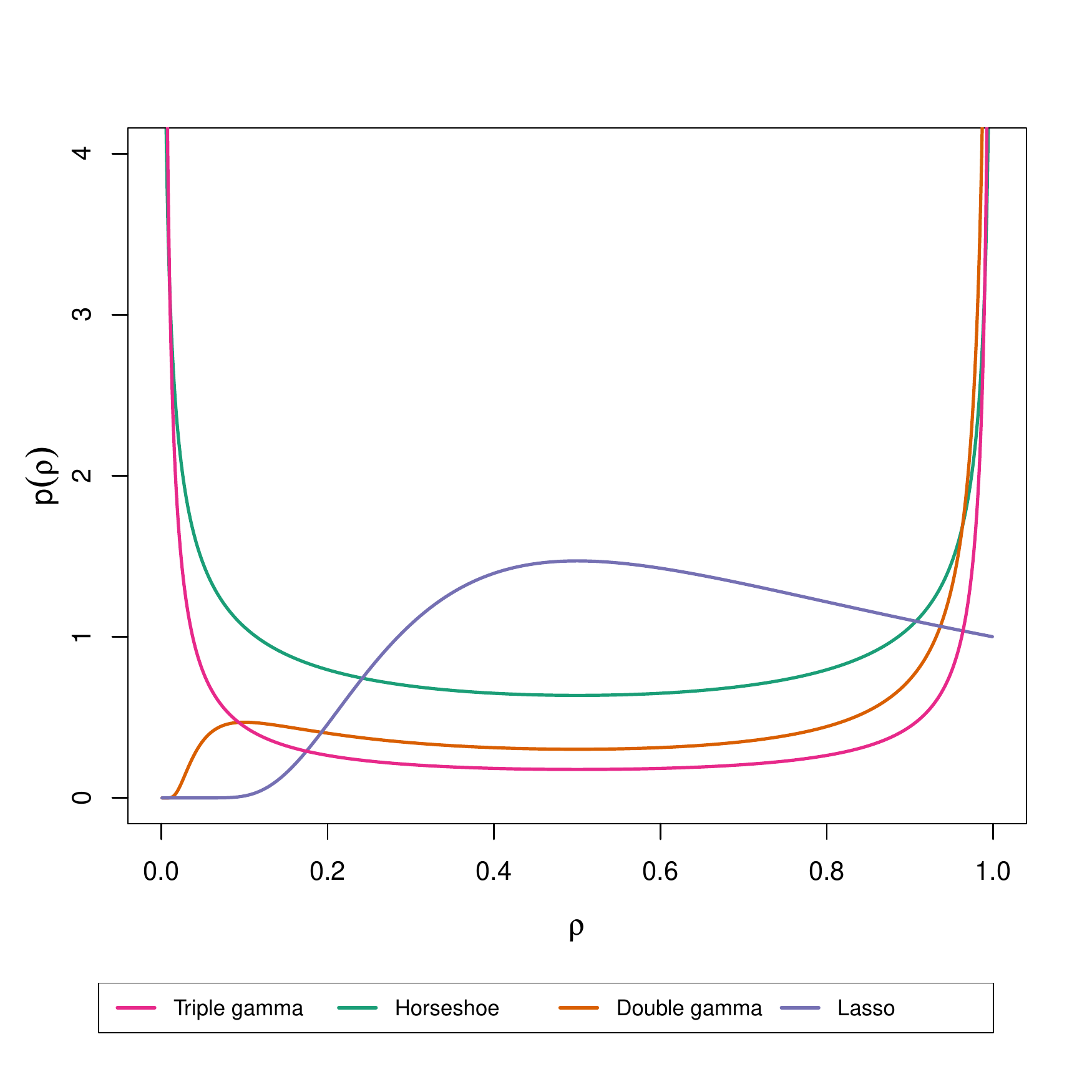}
			\caption{Marginal univariate shrinkage profile  under the triple gamma prior with $a^\xi=c^\xi=0.1$,  in comparison to the Horseshoe prior, the double gamma prior with $a^\xi = 0.1$ the Lasso prior. $\kappa^2_B = 2$ for all the prior specifications.}
			\label{fig:shrink1d}
		\end{figure}

	Also in a variance selection context, it is evident from (\ref{tpbn2}) that  values of  $\kappashr_j\approx 0 $ will introduce no shrinkage on $\theta_j$, whereas
	values of  $\kappashr_j\approx 1 $ will introduce strong shrinkage of $\theta_j$ toward 0.   Hence, the prior  $p(\kappashr_j)$, also called shrinkage
	profile,  will play an instrumental role in the behaviour of different shrinkage priors. Following   \citet{car-etal:hor}, shrinkage priors are often compared in terms of the prior they imply on  $\kappashr_j$,
	i.e. how they handle shrinkage for small \lq\lq observations\rq\rq\ (in our case innovations) and how robust they are
	to large  \lq\lq observations\rq\rq .
	Note that we ideally want a shrinkage profile that has a pole in zero (heavy tails to avoid over-shrinking signals)
	and a pole in one (spikiness to shrink noise).
	The Horseshoe prior, e.g., implies  $\kappashr_j \sim \Betadis{1/2,1/2}$ which is a shrinkage profile that takes this
	much desired form of a   \lq\lq Horseshoe\rq\rq , see Figure~\ref{fig:shrink1d}.

	For the triple gamma prior,  the shrinkage profile is given by the three-parameter beta 
	prior  $p(\kappashr_j)$ provided in  (\ref{densFrho}). For  {$ \phixi=1$},  $\kappashr_j \sim \Betadis{c^\xi,a^\xi}$ and
	{$\kappa_B^2=2 c^\xi/ a^\xi$}.
	Choosing small values $a^\xi<<1$ will  put prior mass close to 1,
	choosing small values  $c^\xi<<1$ will  put prior mass close to 0, whereas  values for both  $a^\xi$   and $c^\xi$   smaller than one will induce
	the  form of a   Horseshoe prior for   $\kappashr_j$.  Evidently, for $ \phixi=1$, a symmetric  triple gamma prior with $a^\xi=c^\xi $
	implies  a   Horseshoe prior for   $\kappashr_j$ that is symmetric around 0.5.
	This is illustrated in  Figure \ref{fig:shrink1d} for a symmetric triple gamma with $a^\xi=c^\xi=0.1$.
	
	In Figure \ref{fig:shrink1d} we can also see the shrinkage profile for the Bayesian Lasso and the double gamma, which correspond to a triple gamma where $c^\xi \rightarrow \infty$. \footnote{Using (\ref{eq:TG}), we obtain the following prior for  $\kappashr_j=1/(1+ \xiF_j)$ by the law of transformation of densities:
		\begin{align*} 
		p(\kappashr_j )=  {\dfrac{1}{\Gamma(a^\xi)}} \left(\frac{a^\xi \kappa_B^2}{2}\right) ^{a^\xi} (1 -\kappashr_j)^{a^\xi -1} {\kappashr_j}^{-(a^\xi +1)}   \exp \left( - \left(\frac{1 -\kappashr_j }{\kappashr_j} \right) \frac{a^\xi \kappa_B^2}{2} \right).
		\end{align*}}
	For the  Bayesian Lasso with $a^\xi=1$
	it is clear that the shrinkage profile $p(\kappashr_j)$ converges to a constant for $\kappashr_j \to 1$, while there is no mass around $\kappashr_j=0$. This means that this prior tends to over-shrink signals, while not shrinking the noise completely to zero. A double gamma prior with $a^\xi < 1$ has the potential to shrink the noise completely to zero,
	as $p(\kappashr_j)$ has a pole at $\kappashr_j = 1$,
	but $p(\kappashr_j)$ has also zero mass around $\kappashr_j=0$, meaning the prior encourages over-shrinking of signals.

		\begin{figure}[t!]
			\centering
			\includegraphics[width=\textwidth]{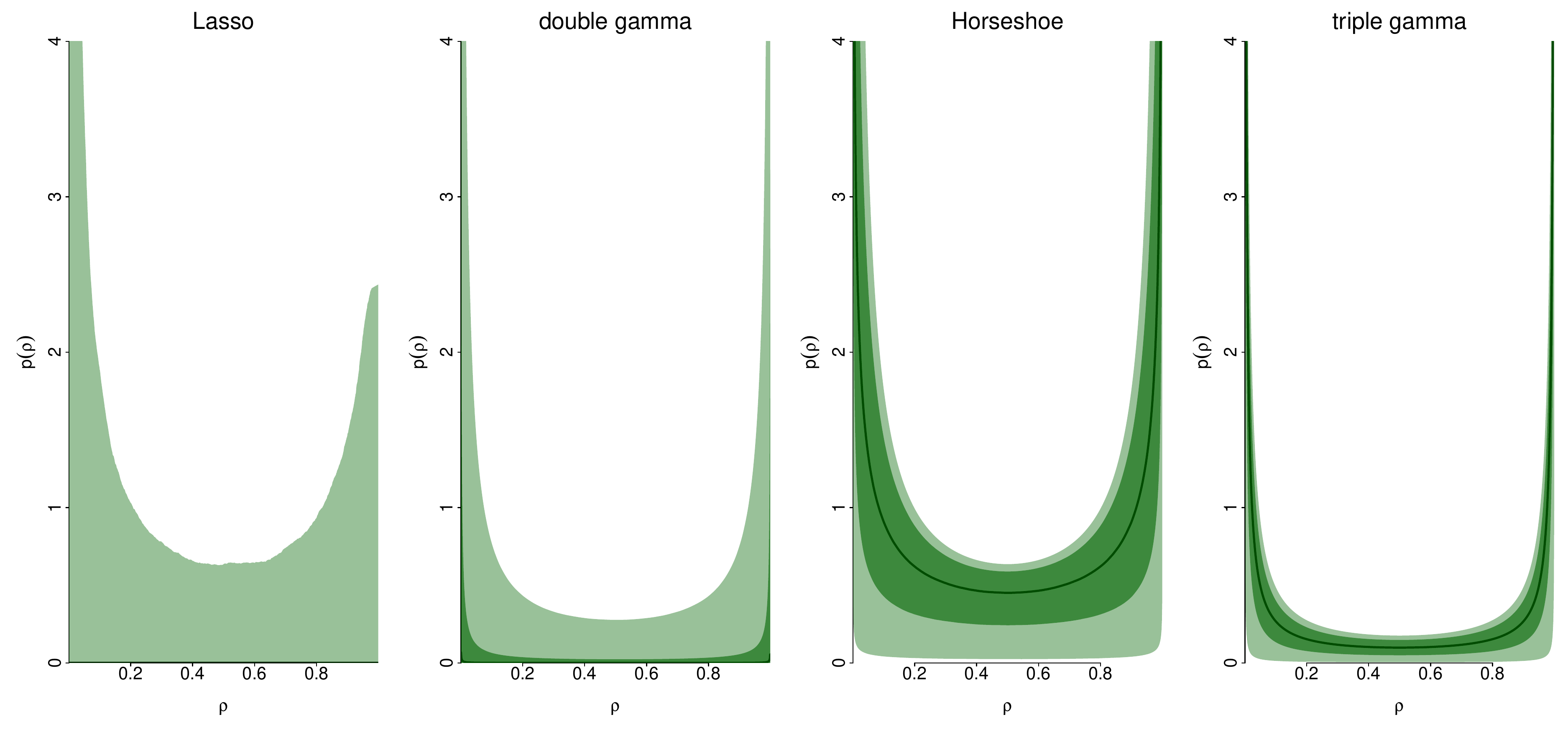}
			\caption{\lq\lq Prior density\rq\rq\ of  shrinkage profiles for (from left to right) Lasso prior, double gamma prior with $a^\xi = 0.2$, Horseshoe prior and triple gamma prior with $a^\xi = c^\xi = 0.1$, when $\kappa_B^2$ is random. The solid line is the median, while the shaded areas represent 50\% and 95 \% prior credible bands.
		We have used $\kappa_B^2 \sim \Gammad{0.01, 0.01}$  for the Lasso and the double gamma, $2/\kappa_B^2 \sim \Fd{1,1}$  for the Horseshoe and  $2/\kappa_B^2 \sim \Fd{0.2, 0.2}$ for the triple gamma.}
			\label{fig:shrink_prior}
		\end{figure}
		
	When we make $\kappa_B^2$ random,  we obtain a \lq\lq prior density\rq\rq\ of  shrinkage profiles,  see Figure \ref{fig:shrink_prior}.
	We can see that such hierarchical versions of the Lasso and the double gamma have shrinkage profiles that resemble the ones of the Horseshoe and the triple gamma.   We have used $\kappa_B^2 \sim \Gammad{0.01, 0.01}$  for the Lasso and the double gamma,
	$2/\kappa_B^2 \sim \Fd{1,1}$  for the Horseshoe and  $2/\kappa_B^2 \sim \Fd{0.2, 0.2}$ for the
	triple gamma, see  {Section~\ref{sec_hyp}}.


	\subsection{BMA-type behaviour}   \label{sec_BMA}

	From the perspective of Bayesian model averaging (BMA),
	an ideal approach for handling sparsity in TVP models would be the use of discrete mixture priors as suggested in \citet{fru-wag:sto},
	\begin{align} \label{eq:DiscreteMix}
	p(\sqrt{\theta}_j) =  (1-\pi) \delta_0 + \pi \cdot  p_{\slab}(\sqrt{\theta}_j),
	\end{align}
	with $ \delta_0$ being a Dirac measure at 0, while $p_{\slab}(\sqrt{\theta}_j)$ is the prior for non-zero variances.
	In terms of shrinkage profiles,  the discrete mixture   prior (\ref{eq:DiscreteMix}) has a spike at $\kappashr_j=1$, with probability  $1-\pi$, and a lot of prior mass at   $\kappashr_j=0$, provided that  the tails of $p_{\slab}(\sqrt{\theta}_j)$ are heavy enough.
	The mixture prior  \eqref{eq:DiscreteMix} is considered the \lq\lq gold standard\rq\rq\ in BMA,
	both theoretically and empirically, see e.g. \citet{joh-sil:nee}.
	However, MCMC inference 
	under this prior is extremely challenging.
	As opposed to this, MCMC inference for the triple gamma prior is straightforward, see Section~\ref{sec_mcmc}.
	
	In this section, we  relate the triple gamma prior to BMA based on the discrete mixture   prior (\ref{eq:DiscreteMix}).
	An interesting insight is that the triple gamma prior  shows a very similar behaviour as  a discrete mixture prior,
	if  both $a^\xi$ and $c^\xi$ approach zero.
	This induces a BMA-type behaviour on the joint shrinkage profile $p(\kappashr_1, \ldots, \kappashr_d)$,
	with a spike at all corner solutions, where some $\kappashr_j$ are very close to one, whereas
	the remaining ones very close to zero.

\begin{figure}[t!]
			\centering
			\includegraphics[width=1\textwidth]{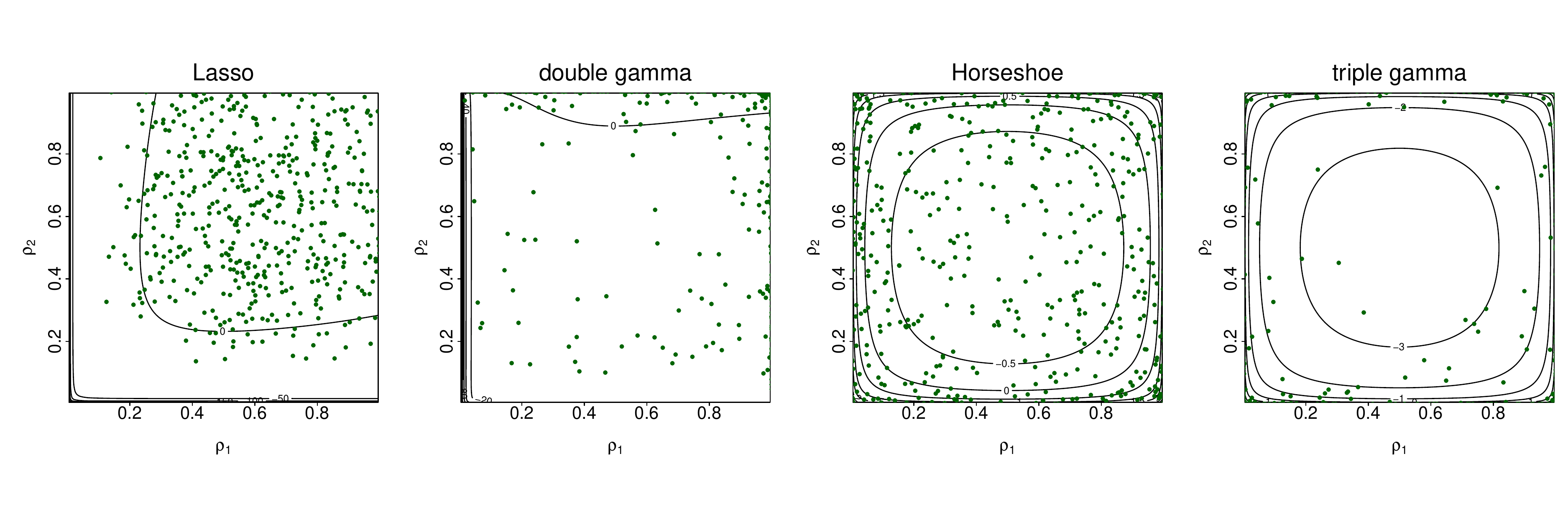}
			\caption{Bivariate shrinkage profile $p(\kappashr_1,\kappashr_2)$
				for (from left to right) the Lasso prior, the double gamma prior with $a^\xi = 0.1$, the Horseshoe prior, and
				the triple gamma prior with $a^\xi=c^\xi=0.1$, with $\kappa^2_B = 2$ for all the priors. The contour plots of the bivariate shrinkage profile are shown, together with 500 samples from the bivariate prior distribution of the shrinkage parameters.}
			\label{fig:shrink_prior_biv}
		\end{figure}

	The bivariate shrinkage profiles shown  in Figure~\ref{fig:shrink_prior_biv}
	give us some intuition about the convergence of a symmetric  triple gamma prior with $a^\xi=c^\xi \rightarrow 0$
	toward  a discrete spike and slab mixture. As opposed to the Lasso and the double gamma prior, the Horseshoe and the
	triple gamma prior put nearly all  prior mass on the \lq\lq corner solutions\rq\rq , which correspond to the four  possibilities
	(a)  $\kappashr_1= \kappashr_2 =0$, i.e. no shrinkage on $\theta_1$ and $\theta_2$,   (b)  $\kappashr_1=1,  \kappashr_2 =0$, i.e. shrinkage of   $\theta_1$ toward 0 and no shrinkage on  $\theta_2$,  (c)  $\kappashr_1=0,  \kappashr_2 =1$, i.e. shrinkage of $\theta_1$ toward 0 and no shrinkage on  $\theta_2$,
	and (d)  $\kappashr_1= \kappashr_2 =1$, i.e. shrinkage  of  both  $\theta_1$ and $\theta_2$  toward 0.

	A very important aspect of BMA is the one of choosing a prior for the model dimension, $K$, see e.g.  \citet{fer-etal:ben} and \citet{ley-ste:eff}. In the discrete mixture prior  (\ref{eq:DiscreteMix}), the distribution  of  $K$ depends on the choice of $\pi$.
	Fixing $\pi$ corresponds to a very informative prior on the model dimension,  for example $\pi=0.5$ assigns more prior probability to models of dimension $d/2$ and lower prior probability to empty or full models.
	In fact, let $\delta_j$ be the indicator that tells us if the $ j$-th coefficient is included in the model, then we have that $K=\sum_{j=1}^d{\delta_j} \sim
	\Bino{d, \pi}$.
	Placing a uniform prior for $\pi$ has been shown to be a good choice, since it corresponds to placing a
prior on $K$ which is uniform on $\{0, \ldots, d\}$.
	Note that $\pi$ will be learned using information from all the variables, that it is a global parameter and will adapt to the degree of sparsity.

	Following  ideas in \cite{car-etal:han}, we believe that a natural way to perform variable selection in the continuous shrinkage prior framework is though thresh-holding. 
	Specifically, we  say that when 
	$ (1-\kappashr_j) > 0.5$, or $\kappashr_j< 0.5$, the variable is included, otherwise it is not.
	Notice that this classification via thresh-holding makes perfectly sense in the case of   a  triple gamma
	of which the Horseshoe is a special case,
	but less so for  a Lasso or double gamma prior, even if the shrinkage profile shows a Horseshoe-like behaviour for  hierarchical versions of these
	priors (see again Figure \ref{fig:shrink_prior}).
	Notice that this implies a prior on the model dimension $K$.
	Specifically,
	\begin{align} \label{dimKpr}
	K = \sum_{j=1}^{d}\indic{ \kappashr_j< 0.5 } \sim
	\Bino{d,    \pi^\xi}, \quad    \pi^\xi= \Prob{\kappashr_j< 0.5 },
	\end{align}
	where  $\kappashr_j| a^\xi, \phi^\xi \sim \TPB{a^\xi, b^\xi, \phi^\xi}$, see (\ref{densFrho}).
	The choice of $\phi^\xi$ (or $\kappa^2_B$) will strongly impact  the prior on  $K$.
	For a symmetric triple gamma with $a^\xi=c^\xi$, for instance, and fixed $\phixi=1$, that is $\kappa^2_B=2$,   we obtain  $K \sim \Bino{d, 0.5}$, since  
	$\pi^\xi=0.5$  regardless of $a^\xi$.
	Hence,  we have to  face similar problems as with fixing  $\pi=0.5$ for  the discrete mixture prior (\ref{eq:DiscreteMix}).
	
	Placing a hyper prior on  $\phitau$  and   $\phixi$ or, equivalently on, $\lambda_B^2$ and  $\kappa^2_B$, as we did in Section~\ref{sec_hyp},
	is as instrumental for  BMA-type variable and variance selection for the triple gamma prior,
	as  is making $\pi$ random  for the discrete mixture prior (\ref{eq:DiscreteMix}).
	Ideally, we would like to have a uniform distribution on the model size $K$.
	We  show in Theorem~\ref{theo2} that the hyperprior  for  $\kappa_B^2 $  defined  in (\ref{hypfinal})
	achieves exactly this goal, since $ \pi^\xi$ is uniformly distributed, see Appendix~\ref{proofs} for a proof.

	\begin{Theorem}\label{theo2}
		For a hierarchical triple gamma prior  with  fixed $a^\xi>0$ and  $c^\xi>0$
		the probability $ \pi^\xi$ defined in (\ref{dimKpr}) follows a uniform distribution, $ \pi^\xi  \sim \Uniform{0,1}$,
		under the   hyper prior
		\begin{align}
		\label{hypfinalproof} \left.  \frac{\kappa_B^2}{2}  \right|  a^\xi , c^\xi   \sim \Fd{2a^\xi ,2 c^\xi}  ,
		\end{align}
		or, equivalently, under  the   hyper prior
		\begin{align}
		\label{hypphiproof} &  \phixi   |  a^\xi , c^\xi   \sim \Betapr{c^\xi,a^\xi }.
		\end{align}
	\end{Theorem}
	

	\section{MCMC algorithm}    \label{sec_mcmc}
	
	Let $\ym=(y_1, \ldots, y_T)$ be the vector of time series observations and let  $\zm$ be the set of all  latent variables and unknown model parameters in a TVP model.
	Moreover, let $\zcond{x}$ denote the set of all unknowns  but $x$.
	Bayesian inference based on MCMC sampling from the posterior $p(\zm|\ym)$ 
	is summarized  in Algorithm~\ref{Algo1}.  The hierarchical priors  introduced in Section~\ref{sec_hyp} are employed,  where
	$(a^\tau,  c^\tau,\lambda_B^2)$ follow   (\ref{priortau}),  $( a^\xi ,  c^\xi)$  follow (\ref{prioraxi}),
	and $\kappa_B^2$  follows (\ref{hypfinal}).
	For certain sampling steps,
	the  hierarchical representation   (\ref{hypkappa}) is used for $\kappa_B^2$,
	and similarly for  $\lambda_B^2$.
	
	Algorithm~\ref{Algo1} extends several existing  algorithms
	such as the  MCMC schemes introduced for the Horseshoe prior by \citet{mak-sch:sim}
	and for the double gamma prior by \citet{bit-fru:ach}.
	We 
	exploit  various representations of the triple gamma prior
	given in Lemma~\ref{Lem1} and choose representation (\ref{TRipleSqphi})
	as the baseline representation of our MCMC algorithm:
	\begin{align*}
	& \beta_j |  \tauphi_j ^2 , \lambdaphi_j^2, \phitau \sim \Normal{0, \phitau \tauphi_j^2/\lambdaphi_j^2},
	\quad  \tauphi_j^2| a^\tau   \sim \Gammad{a^\tau,  1}, \quad \lambdaphi_j^2| c^\tau  \sim \Gammad{c^\tau,  1} ,\\
	& \sqrt{\theta_j} |  \xiphi_j ^2 , \kappaphi_j^2, \phixi \sim \Normal{0, \phixi \xiphi_j^2/\kappaphi_j^2},
	\quad  \xiphi_j^2| a^\xi   \sim \Gammad{a^\xi,  1}, \quad \kappaphi_j^2| c^\xi  \sim \Gammad{c^\xi,  1} ,
	\end{align*}
	where  $\phitau=2 c^\tau/(\lambda_B^2 a^\tau)$ and $\phixi=2 c^\xi/(\kappa_B^2 a^\xi)$.
	All  conditional distributions in our MCMC scheme are available in closed form, expect the ones for $a^\xi$, $c^\xi$, $a^\tau$ and $c^\tau$, for which we will resort to a MH step within Gibbs.
	Several conditional distributions are the same as for the double gamma prior and we apply Algorithm~1 of \citet{bit-fru:ach}. We provide more details on the derivation of the various densities  in Appendix~\ref{sec_details}.

	\begin{alg}{{\bf MCMC inference for TVP models under the triple gamma prior}}\label{Algo1}
		Choose starting values for all global and local shrinkage parameters, i.e.
		$(a^\tau,  c^\tau,\lambda_B^2, a^\xi ,  c^\xi, \kappa_B^2)$ and $\{ \tauphi ^2_j, \lambdaphi_j^2, \xiphi_j^2 , \kappaphi_j^2 \}_{j=1}^d$, and repeat the following steps:
		\begin{itemize} \itemsep 1mm
			\item[(a)]  Define for $j=1, \ldots, d$,
			$\tau^2_j = \phitau\tauphi ^2_j/ \lambdaphi_j^2$ and  $\xi^2_j =  \phixi \xiphi_j^2 / \kappaphi_j^2$
			and sample from the posterior $p(\btildev{0}, \ldots, \btildev{T}, \beta_1, \ldots, \beta_d,
			\sqrt{\theta_1}, \ldots, \sqrt{\theta_d} | \{ \xi_j^2,  \tau_j^2\}_{j=1}^d,\ym)$
			using Algorithm~1, Steps~(a), (b), and (c) in  \citet{bit-fru:ach}.
			In the homoscedastic case, use  Step~(f) of this algorithm to sample from 
			$\sigma^2 | \zcond{\sigma^2}, \ym $.
			For the SV model (\ref{svht}), sample the parameters
			$\mu$, $\phi$, and  $\sigma_\eta^2$  as in \citet{kas-fru:anc}, e.g., using the R-package {\tt stochvol}    \citep{kas:dea}.
			
			\item[(b)]  Use the  prior $p(\sqrt \theta_j | \kappaphi_j^2, a^\xi, c^\xi ) $, marginalized  w.r.t.  $\xiphi_j^2$,
			to sample  $a^\xi$  from  $p(a^\xi| \zcond{a^\xi} , \ym)$ via a random walk MH step on {$ z  = \log (a^\xi /(0.5 -  a^\xi))$. Propose $a \imarg{\xi}{*} = 0.5 \e^{z^*}/(1 + \e^{z^*})$,  where $z^* \sim \Normal{z^{(m-1)}, v^2 }$ and $z^{(m-1)} = \log (a \imarg{\xi}{m-1} /(0.5 -a \imarg{\xi}{m-1} ))$   depends on  the previous value  $a \imarg{\xi}{m-1} $ of $a^\xi$,  accept  $a \imarg{\xi}{*}$}
			with probability 
			\begin{align*}
			\min \left\{1, \frac{ q_a( a\imarg{\xi}{*})}{  q_a(a \imarg{\xi}{m-1} ) } \right\}, \qquad
		    {q_a (a^\xi) = p(a ^\xi  | \zcond{a^\xi}, \ym  ) \, a ^\xi(0.5 - a^\xi)},
			\end{align*}
			and update $\phixi= 2 c^\xi/(\kappa_B^2 a^\xi)$.  Explicit forms for $p(a^\xi| \zcond{a^\xi} , \ym)$ and $\log q_a (a^\xi)$ are provided in (\ref{postaxi}) and (\ref{postaxiq}).
			
			Similarly,   use the  prior $p(\beta_j| \lambdaphi_j^2, a^\tau , c^\tau) $,  marginalized  w.r.t.  to $\tauphi_j^2$,
			to sample  $a^\tau$ via a MH step and update $\phitau=2 c^\tau /( a^\tau \lambda_B^2) $.

			\item[(c)]   Sample $\xiphi_j^2$,  $j=1, \ldots, d$, from a generalized inverse Gaussian distribution,  see  (\ref{GIGxiproof}):
			\begin{align}  \label{GIGxi}
			&  \xiphi_j^2|  \kappaphi_j^2 , \theta_j, a^\xi,\phixi   \sim    \GIG{ a^\xi -\frac{1}{2}, 2, \dfrac{  \kappaphi_j^2  \theta_j }{\phixi } }.
			\end{align}
			Similarly, update $\tauphi_j^2$, $j =1, \ldots, d$, conditional on $a^\tau$:
			\begin{align*} 
			\tauphi_j^2 |   \beta_j,   \lambdaphi_j^2 , a^\tau, \phitau  \sim
			\GIG{a^\tau -\frac{1}{2} , 2, \dfrac{  \lambdaphi_j^2  \beta_j^2 }{\phitau} }.
			\end{align*}
			
			\item[(d)]
			Use  the marginal Student-$t$ distribution $p(\sqrt \theta_j | \xiphi_j^2, c^\xi, \kappa_B^2 )  $ given in (\ref{repSTphi})  to sample  $c^\xi$  from  $p(c^\xi| \zcond{c^\xi} , \ym )$
			via a random walk MH step on {$ z  = \log ({c^\xi}/{(0.5 - c^\xi)})$}.
			Propose {$c \imarg{\xi}{*} = 0.5 \e^{z^*} /( 1 + \e^{z^*})$},  where $z^* \sim \Normal{z^{(m-1)}, v^2 }$ and {$z^{(m-1)} = \log (c \imarg{\xi}{m-1} /(0.5- c \imarg{\xi}{m-1}))$}
			depends on  the previous  value  $c \imarg{\xi}{m-1} $ of $c^\xi$,  accept  $c \imarg{\xi}{*}$
			with probability 
			\begin{align*}
			\min \left\{1, \frac{ q_c( c\imarg{\xi}{*})}{  q_c(c \imarg{\xi}{m-1}) } \right\}, \qquad
			{q _c(c^\xi) = p(c ^\xi  |  \zcond{c^\xi} , \ym ) \, c ^\xi(0.5 - c^\xi)},
			\end{align*}
			and update $\phixi=\frac{2 c^\xi}{\kappa_B^2 a^\xi}$.
			Explicit forms for $p(c^\xi| \zcond{c^\xi} , \ym )$ and $\log q_c (c^\xi)$  are provided in (\ref{postcxi}) and (\ref{logqcxi}).
			
			Similarly, to sample  $c^\tau$ via a MH step use the marginal distribution of $\beta_j| \tauphi_j^2 ,  a^\tau , c^\tau $ with respect to $\lambdaphi_j^2 $
			and update $\phitau=2 c^\tau /( a^\tau \lambda_B^2) $.

			\item[(e)] 
			Sample $\kappaphi_j^2$, for $j=1, \ldots, d$, from following gamma distribution,  see  (\ref{Gkappaproof}):
			\begin{align}  \label{Gkappa}
			\kappaphi_j^2 |   \theta_j,   \xiphi_j^2 ,  c^\xi, \phixi \sim \Gammad{ 1/2 + c^\xi, \dfrac{\theta_j }{2 \phixi \xiphi_j^2}  + 1 }.
			\end{align}
			Similarly, update $\lambdaphi_j^2$, $j =1, \ldots, d$, conditional on $c^\tau$:
			\begin{align*}  
			\lambdaphi_j^2 |  \beta_j,   \tauphi_j^2 , c^\tau,  \phitau  \sim \Gammad{ 1/2 + c^\tau, \dfrac{\beta^2_j}{2 \phitau \tauphi_j^2 }   + 1 }.
			\end{align*}
			
			\item[(f)]  
			Sample $d_2$    from
			$  d_2 | a^\xi, c^\xi, \kappa_B^2  \sim    \Gammad{{a^\xi + c^\xi}, \kappa_B^2 + \frac{2 c^\xi}{ a^\xi}}$,
			see (\ref{d2postproof}); sample from $ \kappa_B^2 $ 
			from following gamma distribution,
			\begin{align} \label{kappapost}
			\kappa_B^2 |  \{\theta_j, \kappaphi^2_j,  \xiphi_j^2\}_{j=1}^d,  a^\xi, c^\xi,  d_2
			\sim
			\Gammad{{\frac{d}{2} + a^\xi}, \dfrac{a^\xi}{4 c^\xi}  \sum_{j=1}^d\frac{ \kappaphi_j^2 }{ \xiphi_j^2 } \theta_j  + d_2   },
			\end{align}
			see (\ref{kappapostproof}),  and update $\phixi=\frac{2 c^\xi}{\kappa_B^2 a^\xi}$.
			
			Similarly, sample $e_2$  from
			$ e_2 | a^\tau, c^\tau,  \lambda_B^2   \sim   \Gammad{{a^\tau + c^\tau}, \lambda_B^2 + \frac{2 c^\tau}{a^\tau} }$,  sample
			$\lambda_B^2$ from
			\begin{align*} 
			&\lambda_B^2|  \{\beta_j, \lambdaphi^2_j,   \tauphi_j^2\}_{j=1}^d,  a^\tau, c^\tau,  e_2   \sim
			\Gammad{ {\frac{d}{2} + a^\tau}, \dfrac{a^\tau}{4 c^\tau}  \sum_{j=1}^d\frac{ \lambdaphi_j^2 }{ \tauphi_j^2 } \beta^2_j  + e_2   },
			\end{align*}
			and update  $\phitau=2 c^\tau /( a^\tau \lambda_B^2) $.

		\end{itemize}
	\end{alg}

	\noindent
	The MCMC scheme in Algorithm~\ref{Algo1} is not a full conditional scheme, as several  steps are based on partially marginalized
	distributions.
	That means that the sampling order matters.
	For instance, in Step~(b), we marginalize w.r.t.   $\xiphi_1^2, \ldots, \xiphi_d^2$,
	hence we need to  update $\xiphi_1^2, \ldots, \xiphi_d^2$ after sampling $a^\xi$, before we update $c^\xi$ in Step~(d) conditional on $\xiphi_1^2, \ldots, \xiphi_d^2$.
	Similarly, due to marginalization  in Step~(d),
	we need to  update $\kappaphi_1^2, \ldots, \kappaphi_d^2$, 
	before we update $d_2$ in Step~(f).
	Furthermore, both Step~(b) and Step~(d) are based on the marginal prior of  $\kappa_B^2$,  given in  (\ref{hypfinal}).
	Hence, in Step~(f),  $d_2$ has to be updated from $d_2 | a^\xi, c^\xi, \kappa_B^2 $, before  $\kappa_B^2$ is updated conditional  on $d_2$.
	
	For a symmetric triple gamma prior, where $a^\xi= c^\xi$, the MCMC scheme in Algorithm~\ref{Algo1} has to be modified only  slightly. Either $q_a ( a^\xi)$ in  Step~(b) is  adjusted and Step~(d) is  skipped,
	setting $c^\xi=a^\xi$,  or    $q_c ( c^\xi)$ in  Step~(d)   is
	adjusted and
	Step~(b) is skipped, setting $a^\xi= c^\xi$.  In Appendix~\ref{sec_details}, we provide details  in (\ref{postaxisym}) for the first case   and  in (\ref{postcxisym}) for the second case. Similar modifications are needed, if  $a^\tau= c^\tau$. All other steps in Algorithm~\ref{Algo1} remain the same for  $a^\xi= c^\xi$ and/or $a^\tau= c^\tau$.


\section{Applications to TVP-VAR-SV models} \label{applications}

\subsection{Model}
Consider an $\MVAR$-dimensional time series, $\bm Y_1, \ldots, \bm Y_T$.
The joint dynamics of such time series can be modeled through a time-varying parameter vector autoregressive model with stochastic volatility (TVP-VAR-SV).
Since the influential paper of \citet{pri:tim} (see \citet{del-pro:tim} for a corrigendum), this model
has become a benchmark for analyzing relationships between macroeconomic variables that evolve over time, see  \citet{nak:tim}, \citet{koo-kor:lar}, \citet{eis-etal:sto}, \citet{cha-eis:bay},
\citet{fel-etal:sop} and \citet{car-etal:lar}, among many others. 
A TVP-VAR-SV model of order $p$ can be expressed as follows:
\begin{align}
\comment{\bm Y_t =  \bm c_t + \Phim_{1,t} \bm Y_{t-1}   +
\Phim_{2,t} \bm Y_{t-2} + \ldots  \Phim_{p,t} \bm Y_{t-p}  +  \errorv_t, \qquad \errorv_t \sim \Normult{\MVAR}{\bm 0, \Sigmam_t}},
\end{align}
where $\bm c_t$ is the $\MVAR$-dimensional intercept,  $\Phim_{j,t}$, for $j = 1, \ldots, p$ is an $\MVAR \times \MVAR$ matrix of time-varying coefficients,  and $\Sigmam_t$ is the time-varying variance covariance matrix of the error term.
The  TVP-VAR-SV model can be written in a more compact notation as
\begin{align}
\bm Y_t = ({\mathbb I}_{\MVAR} \otimes \bm X_t ) \bm \beta_t  + \errorv_t, \qquad \comment{\errorv_t \sim \Normult{\MVAR}{\bm 0, \Sigmam_t} },
\end{align}
where $\bm X_t = ( \bm Y_{t -1}', \ldots,\bm  Y_{t-p}', 1)$ is a row vector of length \comment{$\MVAR p+1$} and
$\bm \beta_t = ({ \bm \beta_{t}^1}', \ldots, { \bm \beta_{t}^\MVAR}')'$, where \comment{$ \bm \beta_{t}^i = ( {\Phim_{1, t}}_{i \bullet}, \ldots, {\Phim_{p, t}}_{i \bullet}, c_{t,i})'$}.
Here, \comment{${\Phim_{j, t}}_{i \bullet}$} denotes the $i$-th row of the matrix \comment{$\Phim_{j, t}$} and $c_{t, i}$ denotes the $i$-th element of $\bm c_t$.

\comment{Following \citet{bit-fru:ach}, we use} an LDLT decomposition of the time-varying covariance matrix, that is $\Sigmam_t = \Amul_t \Dmul_t \Amul_t'$, where $\Dmul_t$ is a diagonal matrix and $\Amul_t$ is lower unitriangular matrix, \comment{see also \cite{car-etal:lar}}. We denote with $a_{ij, t}$ the element at the $i$-th row and $j$-th column of $\Amul_t$, and with $d_{i,t}$ the $i$-th element of the diagonal of $\Dmul_t$. In total, we have $\MVAR(\MVAR-1)/2 + \MVAR(\MVAR p+1)$ (potentially) time-varying parameters.
Using the LDLT decomposition, we can rewrite the system as:
\begin{align*}
\bm Y_t = ({\mathbb I}_{\MVAR} \otimes X_t ) \bm \beta_t  + \comment{ \Amul_t   \etav_t, \qquad  \etav_t \sim
\Normult{\MVAR}{\bm 0, \Dmul_t}} .
\end{align*}
\noindent
This, in turn, allows us to write
\begin{align*}
y_{1,t} =& \bm X_t \bm \beta_{t}^1 + \eta_{1,t},  \quad \eta_{1,t} \sim \comment{\Normal{0, d_{1,t}} }\\
y_{2,t} =& \bm X_t \bm \beta_{t}^2 + a_{21,t} \eta_{1,t} + \eta_{2,t}
,  \quad \eta_{2,t} \sim \comment{\Normal{0, d_{2,t}} }\\
y_{3,t} =& \bm  X_t \bm \beta_{t}^3 + a_{31,t} \eta_{1,t} +  a_{32,t}\eta_{2,t} + \eta_{3,t},  \quad \eta_{3,t} \sim \comment{\Normal{0, d_{3,t}} }\\
& \qquad \vdots
\end{align*}
\noindent
Generalizing, for the $i$-th equation we have that
\begin{align*}
y_{i,t} = \bm X_t \bm \beta_{t}^i + \sum_{j=1}^{i-1} a_{ij, t} \eta_{j,t} + \eta_{i,t} ,   \quad \eta_{i,t} \sim \comment{\Normal{0, d_{i,t}} },
\end{align*}
with independent error terms $\eta_{i,t}$ across equations.
In practice, the $i$-th equation of the system  can be written as a TVP regression model where the residuals of the preceding $i - 1$ equations have been added as "regressors".
 The time-varying regression parameters are assumed to follow a random walk, specifically
\begin{align*}
&\beta_{j, t}^i =  \comment{\beta_{j, t-1}^i}  + v_{ij,t}, \quad   v_{ij,t} \sim \comment{\Normal{0, \theta^\beta_{ij}}}, \quad \text{for}\quad  i = 1, \ldots, \MVAR, \quad  \text{and} \quad j   = 1, \ldots, \comment{\MVAR p+1}, \\
&a_{ij,t} =   a_{ij,t-1} + w_{ij, t}, \quad  w_{ij, t} \sim \comment{\Normal{0, \theta^a_{ij}}},  \quad \text{for}\quad  i = 1, \ldots, \MVAR , \quad \text{and} \quad j = 1, \ldots, i-1.
\end{align*}
\comment{with initial values $\beta_{j, 0}^i \sim \Normal{\beta^\beta_{ij}, \theta^\beta_{ij}}$ and $a_{ij,0} \sim \Normal{\beta^a_{ij}, \theta^a_{ij}}$}. Here, $\beta_{j, t}^i$ denotes the $j$th element of the vector $\bm \beta_t^i$.
\comment{Shrinkage priors are then employed row wise, for the initial expectations $\beta^\beta_{ij}$ and  $\beta^a_{ij}$ as well as the variances $\theta^\beta_{ij}$ and $\theta^a_{ij}$.
To allow for greater flexibility in the prior structure, the
$\beta^\beta_{ij}$ and $\beta^a_{ij}$ are assumed to follow independent shrinkage priors, and similarly for
$\theta^\beta_{ij}$ and $\theta^a_{ij}$:
\begin{align}  \label{priorlocal}
	& \beta^{\xdum}_{ij} \sim \Normal{0, \phitaui{\xdum}{i} \tauij{\xdum}{ij} /\lambdaij{\xdum}{ij}},	\quad  \tauij{\xdum}{ij} \sim \Gammad{\atau{\xdum}{i},  1}, \quad \lambdaij{\xdum}{ij}  \sim \Gammad{\ctau{\xdum}{i},  1} ,
\quad \phitaui{\xdum}{i}=2 \ctau{\xdum}{i}/(\lambdai{\xdum}{i} \atau{\xdum}{i} ) ,\\
	& \sqrt \theta^{\xdum}_{ij}  \sim \Normal{0, \phixii{\xdum}{i} \xiij{\xdum}{ij} /\kappaij{\xdum}{ij}},	\quad  \xiij{\xdum}{ij}  \sim \Gammad{\axi{\xdum}{i},  1}, \quad \kappaij{\xdum}{ij} \sim \Gammad{\cxi{\xdum}{i},  1} ,
\quad \phixii{\xdum}{i}=2 \cxi{\xdum}{i}/(\kappai{\xdum}{i} \axi{\xdum}{i} ), \nonumber
\end{align}
where $\xdum=\beta$ for the VAR-coefficients and  $\xdum=a$ for the elements of $\Amul$.
Following Section~\ref{sec_hyp}, the priors for the global shrinkage parameters read
\begin{align} \label{priorkappa}
& \lambdai{\xdum}{i}| \atau{\xdum}{i}, \ctau{\xdum}{i} \sim \Fd{2 \atau{\xdum}{i}, 2 \ctau{\xdum}{i}},  \quad
 2\atau{\xdum}{i} \sim \Betadis{\alpha_{a^\tau},  \beta_{a^\tau}} ,
	\quad 2 \ctau{\xdum}{i} \sim \Betadis{\alpha_{c^\tau},  \beta_{c^\tau}}, &\\
& \kappai{\xdum}{i} | \axi{\xdum}{i}, 2 \cxi{\xdum}{i} \sim \Fd{2 \axi{\xdum}{i}, 2 \cxi{\xdum}{i}}, \quad 	
2\axi{\xdum}{i} \sim \Betadis{\alpha_{a^\xi},  \beta_{a^\xi}} ,
	\quad 2 \cxi{\xdum}{i} \sim \Betadis{\alpha_{c^\xi},  \beta_{c^\xi}}.& \nonumber
\end{align}}
Finally, we assume that the \comment{idiosyncratic shocks $\eta_{i,t} \sim \comment{\Normal{0, d_{i,t}} }$ follow an SV model as in (\ref{svht}), with row specific parameters}. Specifically, let $h_{i,t} = \log d_{i, t} $, we have that the logartihm of the elements of the diagonal matrix $\Dmul$ follow \comment{independent} AR(1) processes:
\begin{align*}
h_{i, t} =  \mu_i + \comment{\phi_i} (h_{i,t-1} - \mu_i) + \nu_{i,t}, \quad \nu_{i,t} \sim \comment{\Normal{0, \sigma_{\eta,i}^2}},
\end{align*}
for $i = 1, \ldots, \MVAR$. Here, $\mu_i$ is the mean of the $i$th log-volatility, \comment{$\phi_i$} is the equation specific persistence parameter, and \comment{$\sigma_{\eta,i}^2$} is the variance of the $i$th log-volatility.

\subsection{A brief sketch of the TVP-VAR-SV MCMC algorithm}

Our algorithm exploits the aforementioned unitriangular decomposition to estimate the model parameters equation-by-equation. \comment{Due to the prior structure introduced in (\ref{priorlocal}),} the estimation of the \comment{$\betav_{t}^i$} and the $a_{ij, t}$s is separated into two blocks, with the algorithm cycling through the equations, alternating between \comment{sampling  $\betav_{t}^i$ conditional on $\Sigmam_t$ and sampling the $a_{ij, t}$s and $d_{i,t}$s conditional on the VAR coefficients $\betav_{t}^i$}.  Given a set of initial values, the algorithm repeats the following steps:

\begin{alg}
{\bf MCMC inference for TVP-VAR-SV models under the triple gamma prior} Choose starting values for \comment{all global and local shrinkage parameters in prior (\ref{priorlocal}) for each equation}
and repeat the following steps:
\begin{itemize} \itemsep 1mm
	\item[] For $i = 1, \ldots, \MVAR$: 
	\begin{enumerate} \itemsep 1mm
	\item[\comment{(a)}] Conditional on \comment{$\Amul_t$ and $\Dmul_t$}, create $\check y_{i,t} =  y_{i, t} - \sum_{j=1}^{i-1} a_{ij, t} \comment{\eta_{j,t}}$ and use Algorithm \ref{Algo1} (sans the step for the variance of the observation equation) on the TVP regression $$\check y_{i,t} =  \bm X_t \bm \beta_{t}^i + \eta_{i,t}, \quad \comment{\eta_{i,t} \sim \comment{\Normal{0, d_{i,t}} }},$$ to draw from the conditional posterior distribution of \comment{the time-varying VAR-coeffcients in row $i$,} $\bm \beta_{t}^i,$ for $t = \comment{0}, \ldots, T,$ \comment{their initial expectations and process variances $\beta^{\beta}_{ij}$ and  $\theta^{\beta}_{ij}$,  as well as their local and global shrinkage  parameters
$\tauij{\beta}{ij}, \lambdaij{\beta}{ij}, \xiij{\beta}{ij}, \kappaij{\beta}{ij},
\lambdai{\beta}{i}, \kappai{\beta}{i} , \atau{\beta}{i}, \ctau{\beta}{i},
 \axi{\beta}{i}$, and $\cxi{\beta}{i}$.}

	\item[\comment{(b)}] For $i>1$, create $ y_{i,t}^\star=  y_{i, t} - \bm X_t
	\bm \beta_{t}^i$, conditional on $\bm \beta_{t}^i$, and again use Algorithm \ref{Algo1} on the TVP regression $$y_{i,t}^\star= \sum_{j=1}^{i-1} a_{ij, t} \eta_{j,t}  + \eta_{i,t}, \quad \comment{\eta_{i,t} \sim \comment{\Normal{0, d_{i,t}} }}, $$
(where the residuals from the previous $i - 1$ equations are used as regressors) to sample  \comment{the volatilities $d_{i,t}$ and the time-varying coefficients of $\Amul_t$ in row $i$, $a_{ij, t}$, for $t = 0, \ldots, T$ from the respective conditional posteriors, as well as  the  initial expectations and process variances $\beta^{a}_{ij}$ and  $\theta^{a}_{ij}$ and the local and global shrinkage  parameters
$\tauij{a}{ij}, \lambdaij{a}{ij}, \xiij{a}{ij}, \kappaij{a}{ij},
\lambdai{a}{i}, \kappai{a}{i} , \atau{a}{i}, \ctau{a}{i},
 \axi{a}{i}$, and $\cxi{a}{i}$.}
\end{enumerate}
\end{itemize}
\end{alg}
\noindent In the following applications, we run our algorithm for $M = 200000$ iterations, discarding the first $100000$ iterations as burn-in, and then keeping the output of one every $100$ iterations.

\subsection{Illustrative example with simulated data} \label{sec:example}
To illustrate the merit of our methodology in the context of TVP-VAR-SVs, we simulate data from two TVP-VAR-SVs with $T = 200$ points in time, $p = 1$ lags and $\MVAR = 7$ equations, with varying degrees of sparsity. In the dense regime, approximately 30\% of the values of $\bm \beta$ and $\bm \theta$ (here referring to the means of the initial states and the variances of the innovations as defined in Section \ref{sec:triplegamma}, respectively) are truly zero, while in the sparse regime approximately 90\% are truly zero.
We show results for the triple gamma prior, the Horseshoe prior, the double gamma and the Lasso.

Regarding the priors on the hyperparameters, \comment{we use 
  prior (\ref{priorkappa}) with  $\alpha_{a^\tau}=\alpha_{c^\tau} = \alpha_{a^\xi}= \alpha_{c^\xi}=1$ and
$\beta_{a^\tau}=\beta_{c^\tau}=\beta_{a^\xi}=\beta_{c^\xi}=6$  for the triple gamma.}
The probability density function of the corresponding beta prior is monotonically increasing, with a maximum at $0.5$. This prior places positive mass in a neighborhood of the Horseshoe, but allows for more flexibility. In practice, placing a prior on the spike and slab parameters of the triple gamma, instead of fixing them to 0.5 as in the Horseshoe, allows us to learn the shrinkage profile from the data. Moreover, since the spike and the slab parameters are allowed to be different, the shrinkage profile can be asymmetric and adapt to the sparseness of the data.

\comment{We assume  that the global shrinkage parameters $\lambdai{\beta}{i},\kappai{\beta}{i}, \lambdai{a}{i}$, and $\kappai{a}{i}$ follow  a $\Fd{1,1}$ distribution for the Horseshoe prior  which corresponds to the prior in \cite{car-etal:han} and   a $\Gammad{0.001, 0.001}$  distribution for the Lasso and the double gamma prior, as  suggested in \cite{bel-etal:hie_tv} and \cite{bit-fru:ach}.}
\comment{Concerning  the spike parameters $\atau{a}{i}, \axi{a}{i}, \atau{\beta}{i}$, and $\axi{\beta}{i}$ of the double gamma, we employ a rescaled beta prior 
 to force them to be smaller than $0.5$. Specifically, we use a  $B (4 , 6)$  prior which places most of its mass between $0.05$ and $0.4$}, a range that \cite{bit-fru:ach} have found to induce desirable shrinkage characteristics.


\begin{figure}[t!]
	\centering
	\hspace*{-1cm}
	\includegraphics{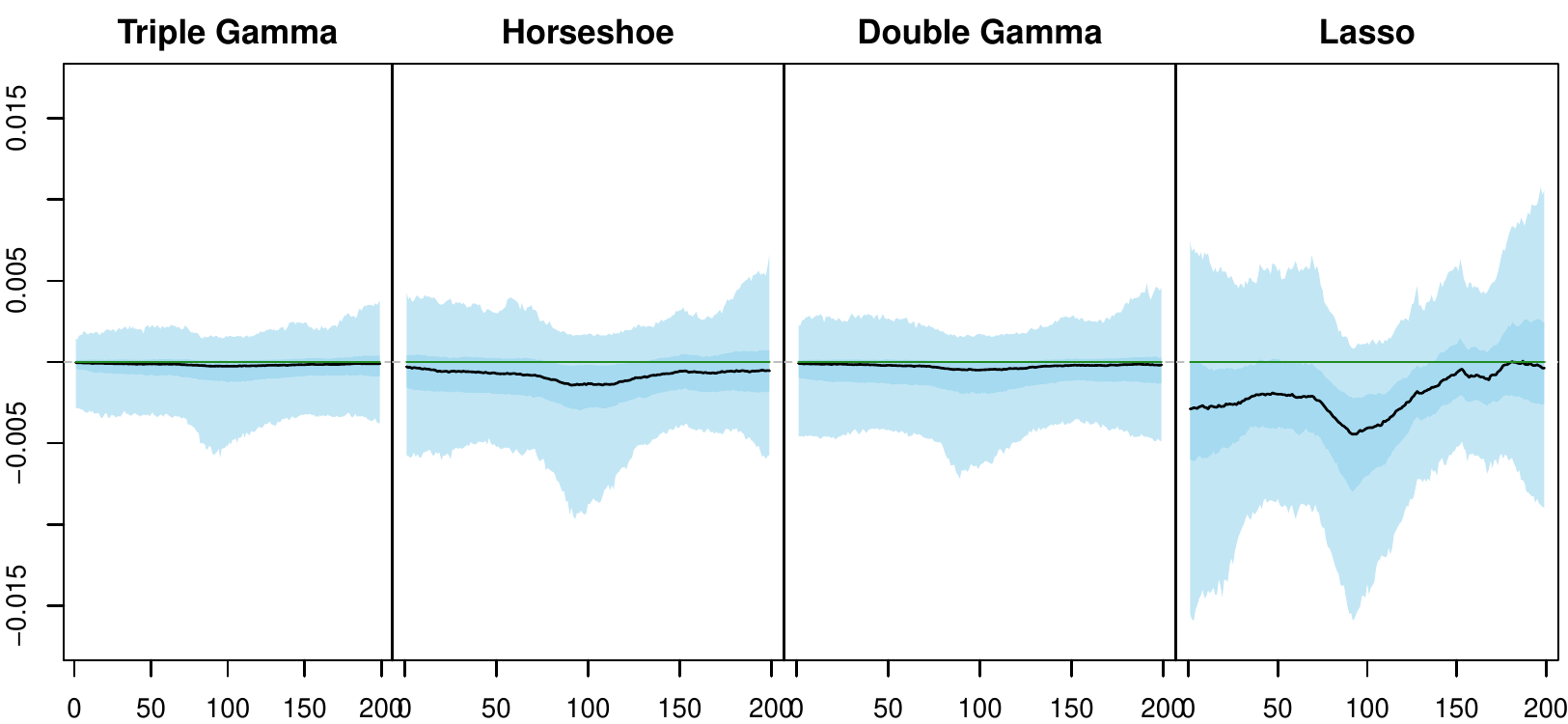}
	\caption{Posterior path against time for a constant non-significant parameter in the sparse regime.}
	\label{fig:statesparsestat}
\end{figure}

Figure \ref{fig:statesparsestat} shows the posterior path against time for a constant non-significant parameter, that is one for which $\theta_{ij}^\beta = 0$ and $\beta_{ij}^\beta = 0$ for all times, in the sparse regime.
The entire set of states for the triple gamma prior can be found in Appendix \ref{sec:simulated_app}.
Note that, while the zero line is contained in the 95\% posterior credible interval for all priors, said interval is thinner under the triple gamma prior and the double gamma prior than under the Lasso and the Horseshoe prior. However, the light tails of the double gamma prior, as the ones of the Lasso, can over-shrink weak signals.

\begin{figure}[t!]
	\centering
	\hspace{-2.5cm}
	\begin{minipage}{0.45\textwidth}
		\includegraphics[width=1.2\textwidth]{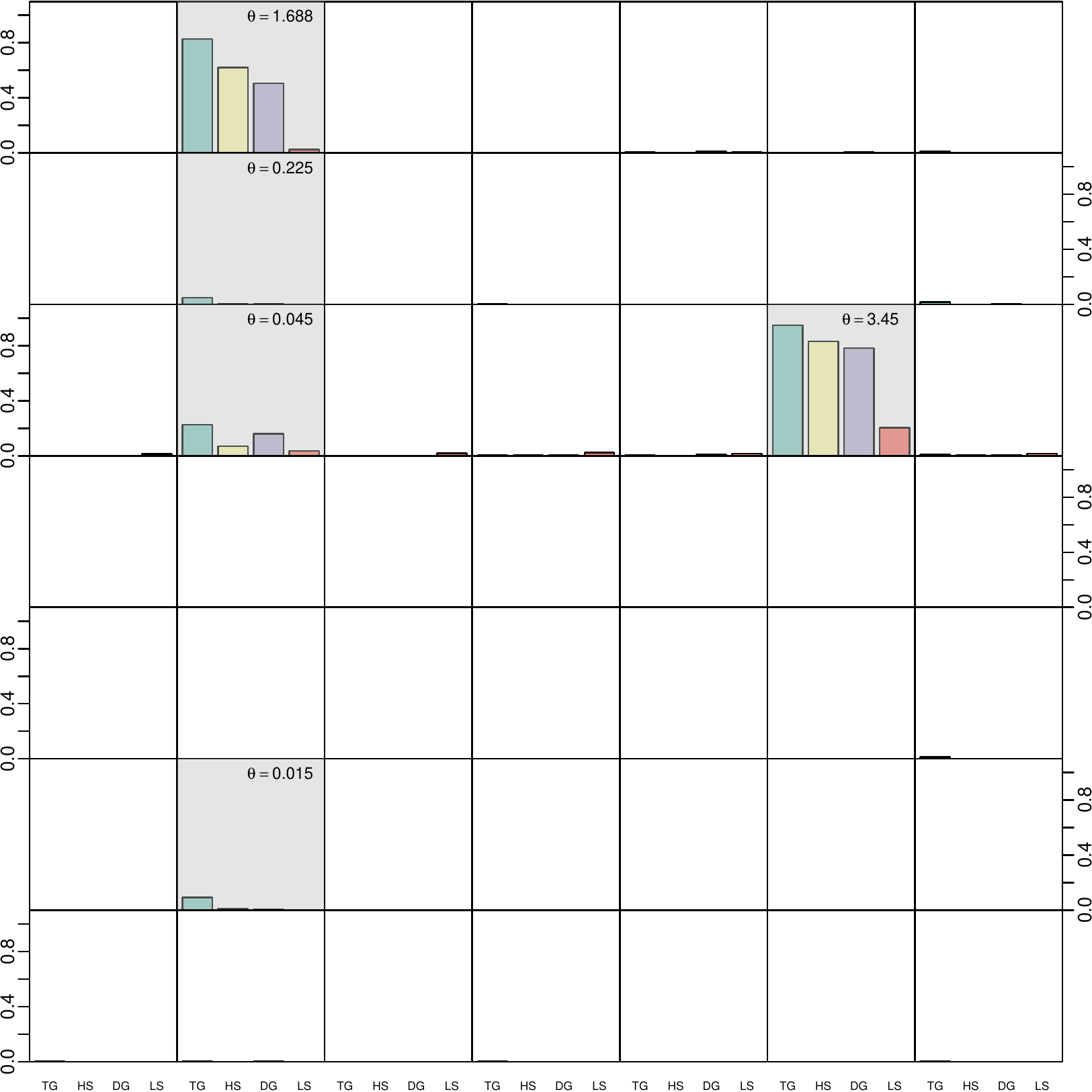}
	\end{minipage}
	\hspace{1.3cm}
	\begin{minipage}{0.45\textwidth}
		\includegraphics[width=1.2\textwidth]{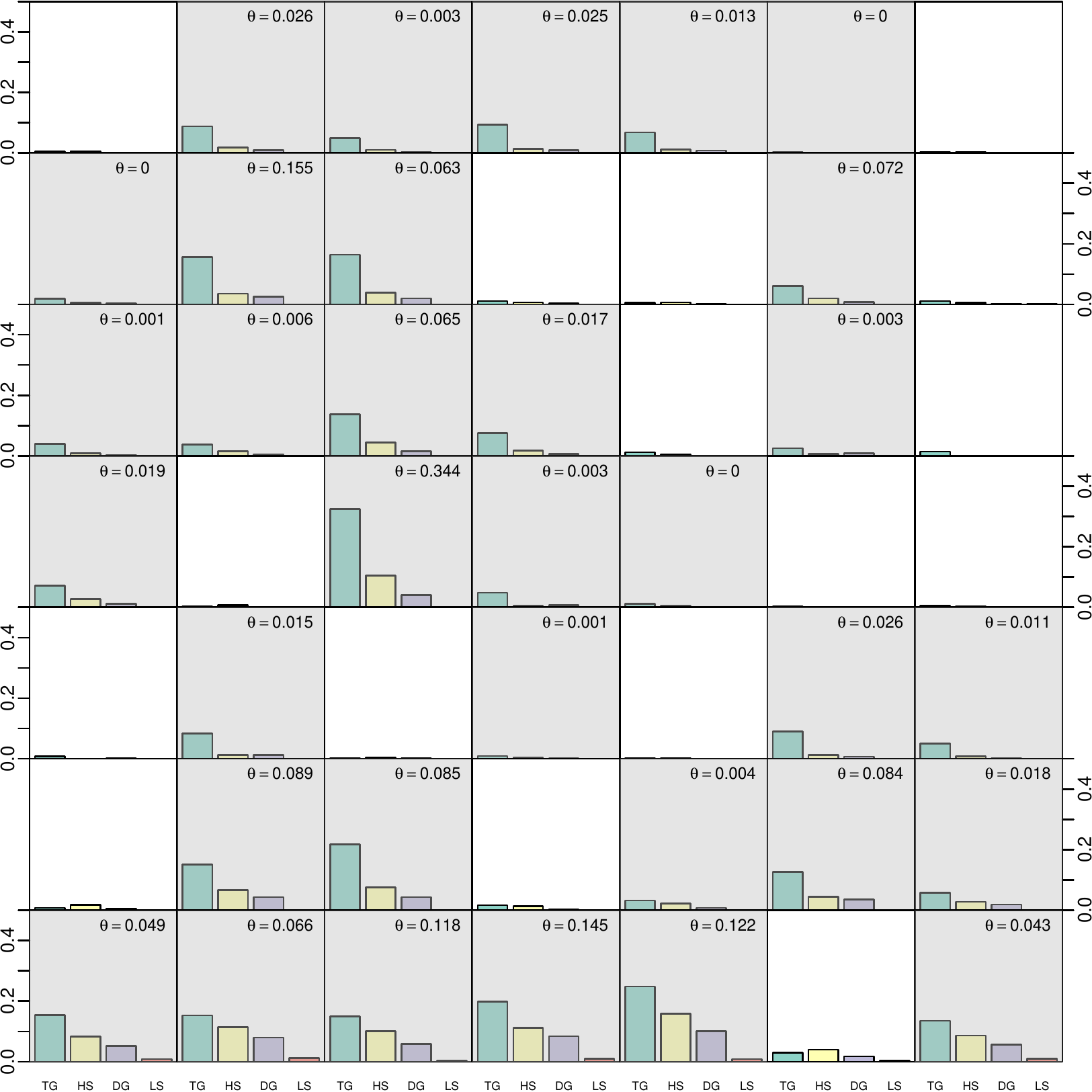}
	\end{minipage}
	\caption{Posterior inclusion probability for the $\theta^\beta_{ij}$'s in the sparse and dense regime, under the triple gamma prior, the Horseshoe prior, the Lasso prior and the double gamma prior. The true values of the $\theta^\beta_{ij}$'s are reported in each cell.}
	\label{fig:incl_synth}
\end{figure}

The above statement becomes clearer when looking at the posterior inclusion probabilities. We calculate the posterior inclusion probabilities based on the thresholding approach introduced in Section \ref{sec_BMA}, comparing the fully unimpeded triple gamma prior to widely used special cases.
Figure \ref{fig:incl_synth}  show the posterior inclusion probability for the variance of the innovations ($\theta^\beta_{ij}$'s) under four different shrinkage priors, for the sparse and the dense scenario, respectively. The cells are shaded in gray when the corresponding true state parameter is time-varying ($\theta^\beta_{ij} \neq 0 $), while the background is white when the corresponding true state parameter is not time-varying ($\theta^\beta_{ij}  = 0$).
The posterior inclusion probabilities under the triple gamma prior are consistently higher for the states for which the parameters are actually time-varying.
In some cases, that heavier tails of our prior pick up signals that the other shrinkage priors are not able to capture. On the other hand, the triple gamma identifies constant parameters, effectively pulling the appropriate $\theta^\beta_{ij}$'s to zero.

\subsection{Modeling area macroeconomic and financial variables in the Euro Area}
\label{sec:application}
Our application investigates a subset of the area wide model of the European Union of \cite{RePEc:eee:ecmode:v:22:y:2005:i:1:p:39-59}, which comprises quarterly macroeconomic data spanning from 1970 to 2017. We include 7 of the variables present in the dataset, namely real output (YER), prices (YED) , short-term interest rate (STN), investment (ITR), consumption (PCR),  exchange rate (EEN) and unemployment (URX). A more detailed description of the data and the transformations performed to make the time series stationary can be found in Table \ref{tabdata} in Appendix \ref{sec:application_app}. To stay in line with the literature, e.g. \cite{feldkircher2017sophisticated}, we estimate a TVP-VAR-SV model with $p = 2$ lags on all endogenous variables. The hyperparameter choices are the same as in Section \ref{sec:example}. As in the example with simulated data, we run the algorithm for $M = 200000$ iterations, discarding the first $100000$ iterations as burn-in, and then keeping the output of one every $100$ iterations.

\begin{figure}[t!]
	\centering
	\hspace{-2.5cm}
	\begin{minipage}{0.45\textwidth}
		\includegraphics[width=1.3\textwidth]{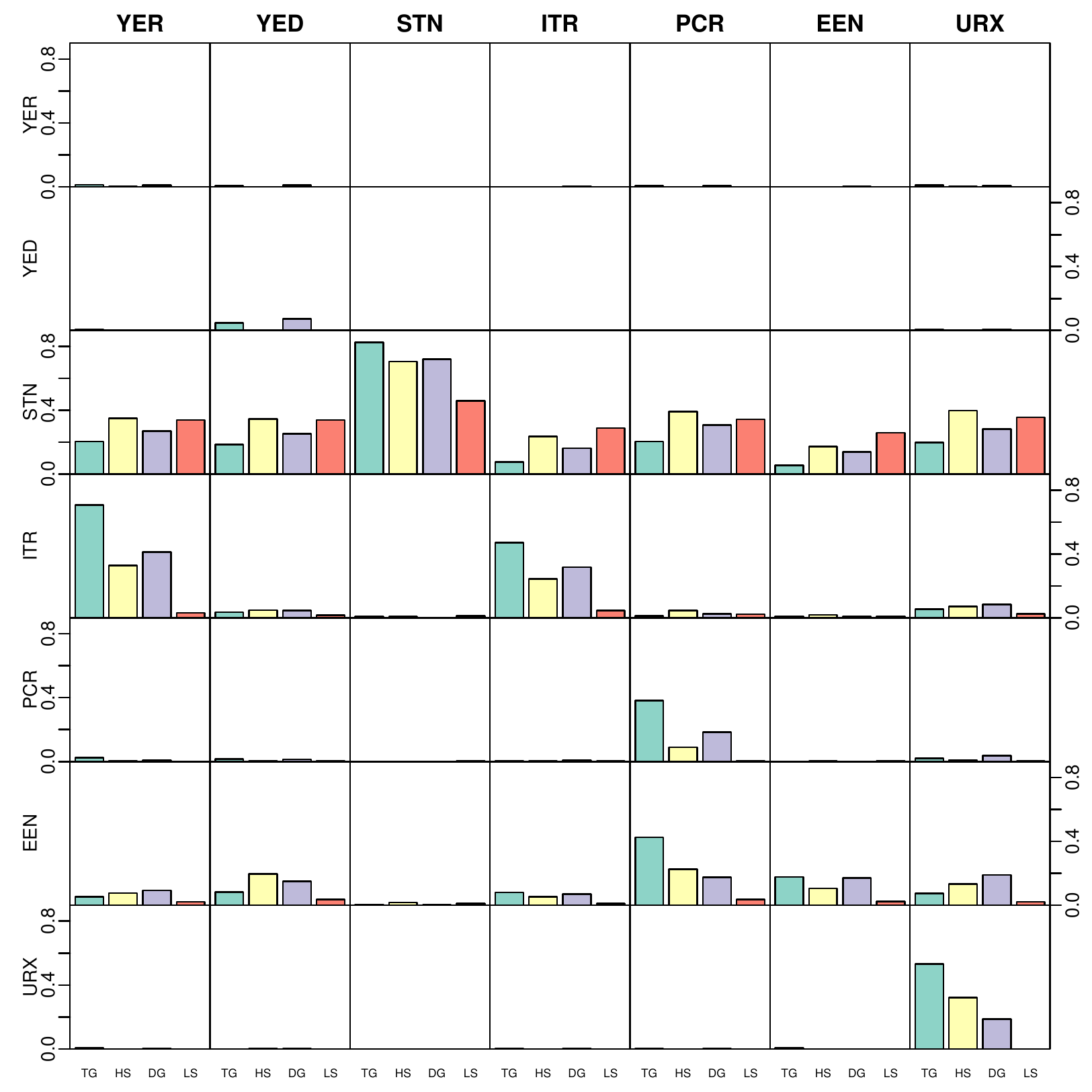}
	\end{minipage}
	\hspace{1.8cm}
	\begin{minipage}{0.45\textwidth}
		\includegraphics[width=1.3\textwidth]{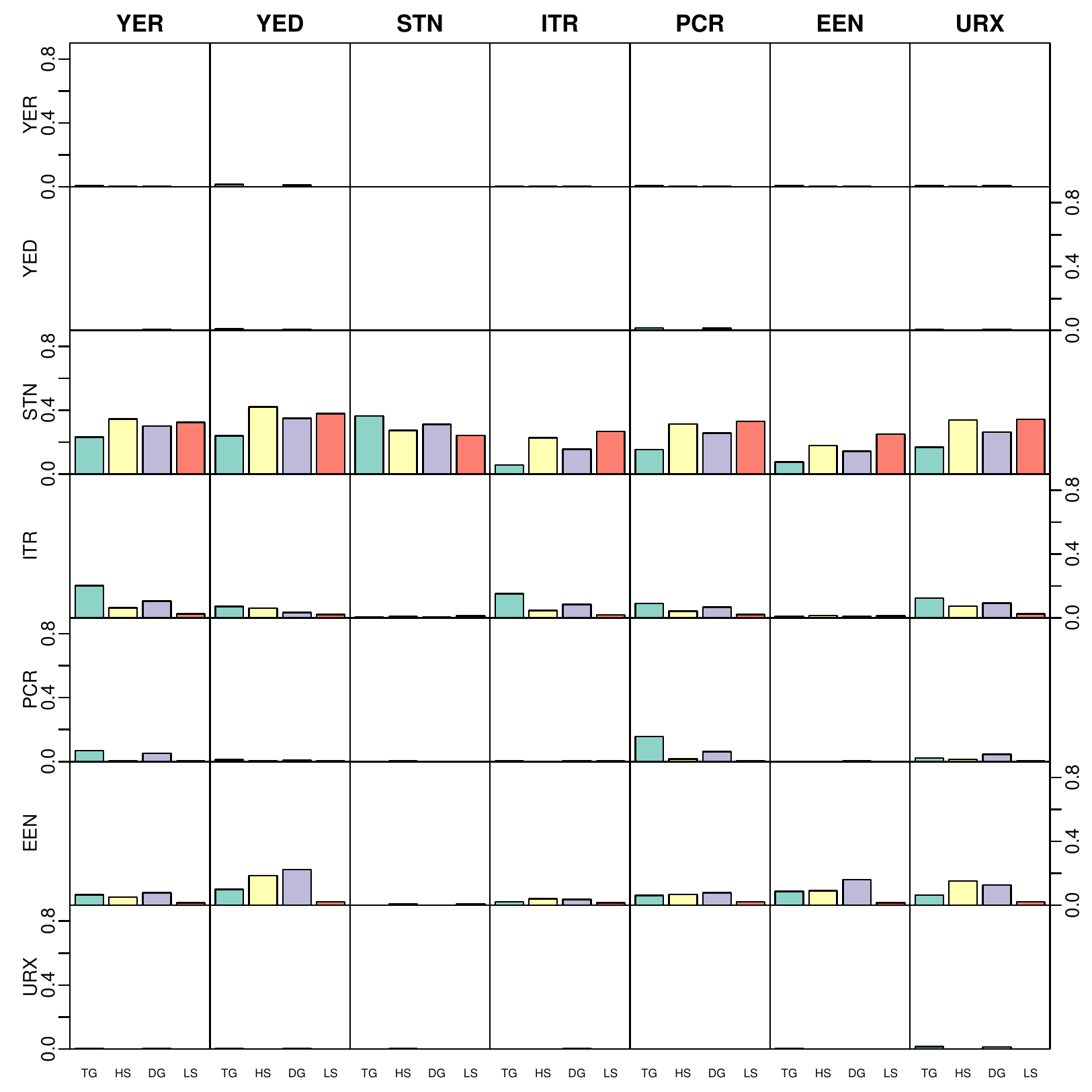}
	\end{minipage}
	\caption{Posterior inclusion probability for state parameters \comment{$\beta_{ij}^\beta$} associated with the first lag (on the left) and with the second lag (on the right), for the EA data under the triple gamma prior, the Horseshoe prior, the Lasso prior and the double gamma prior.}
	\label{fig:incl_beta_EA}
\end{figure}

\begin{figure}[t!]
	\centering
	\hspace{-2.5cm}
	\begin{minipage}{0.45\textwidth}
		\includegraphics[width=1.3\textwidth]{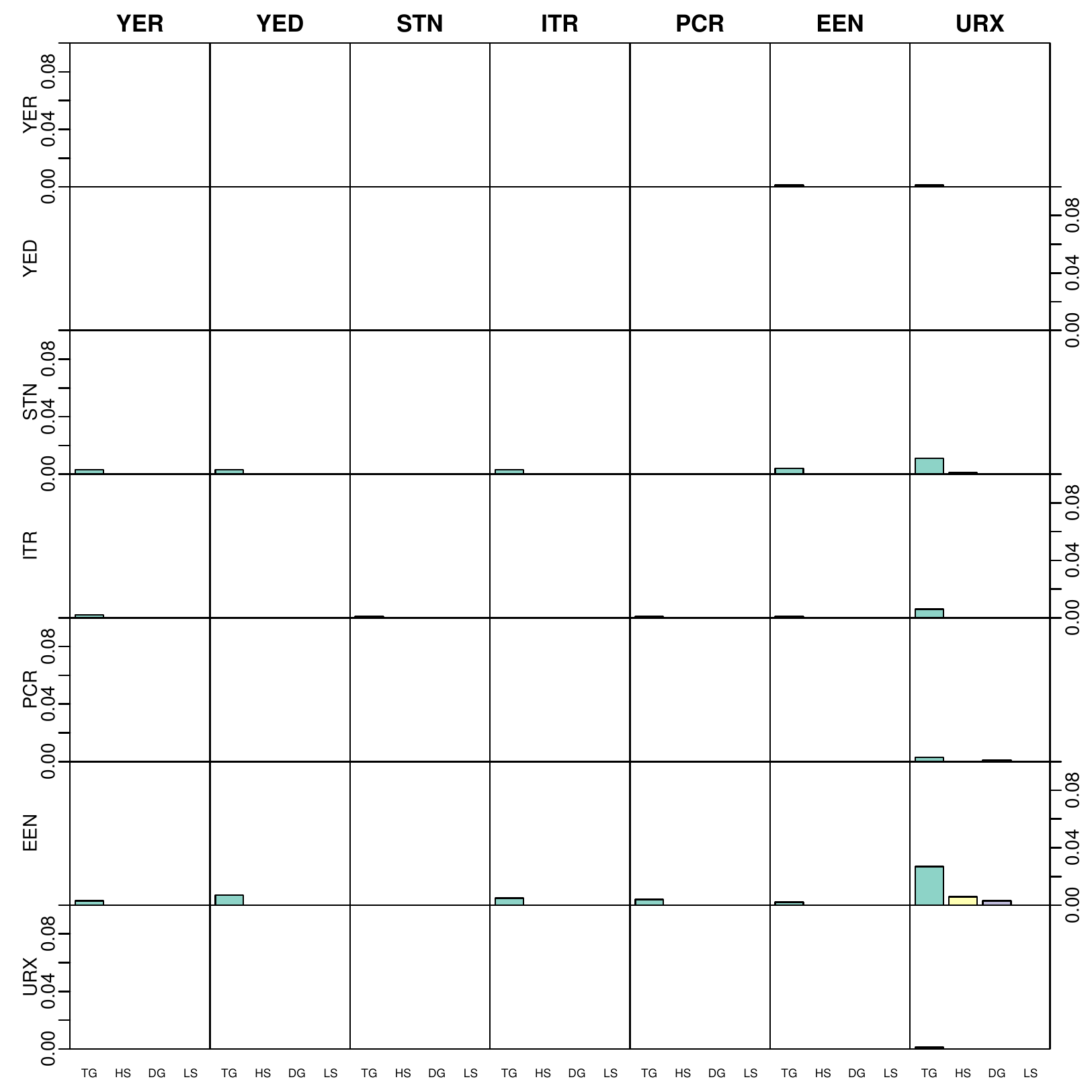}
	\end{minipage}
	\hspace{1.8cm}
	\begin{minipage}{0.45\textwidth}
		\includegraphics[width=1.3\textwidth]{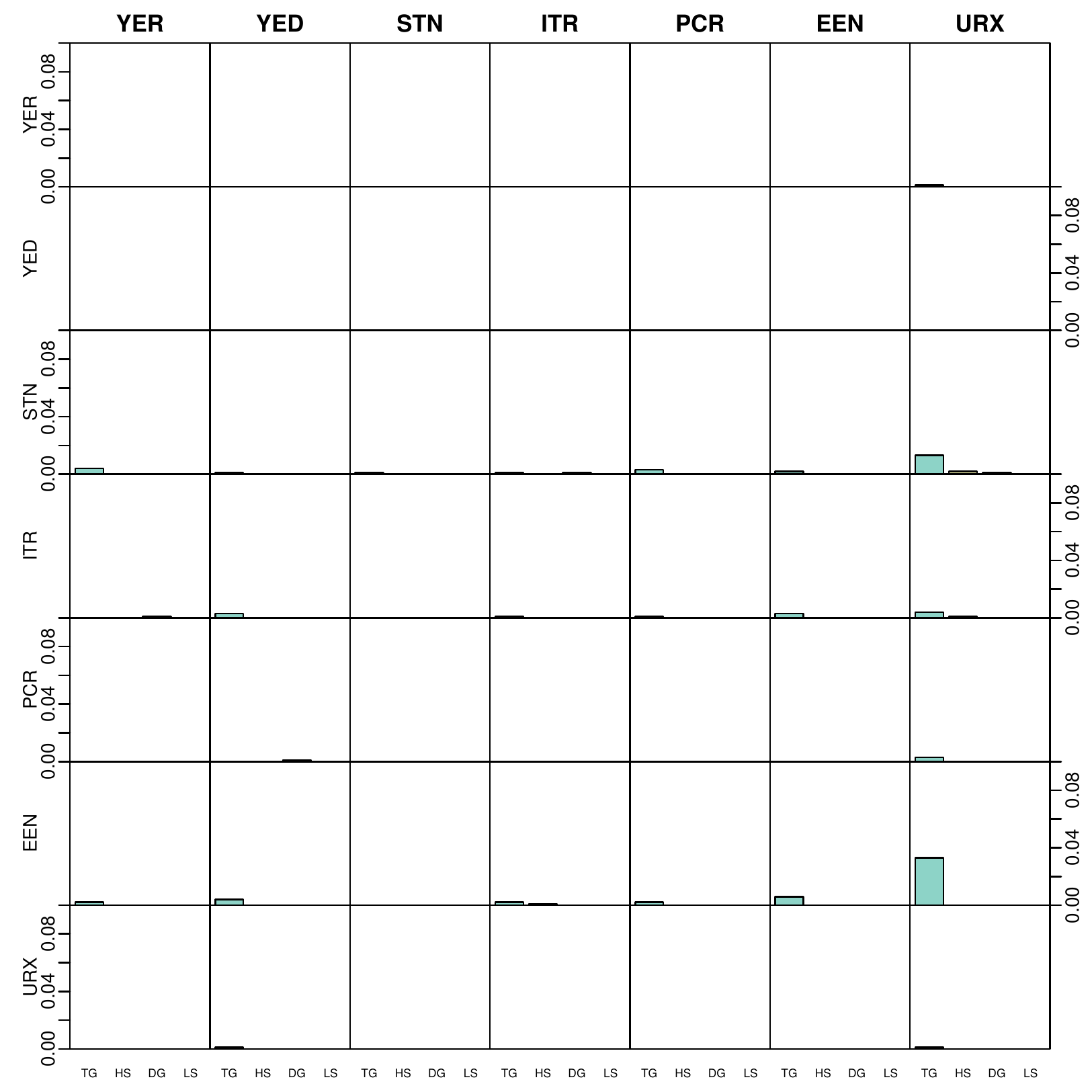}
	\end{minipage}
	\caption{Posterior inclusion probability for $\theta^\beta_{ij}$'s associated with the first lag on the left and with the second lag on the right, for the EA data under the triple gamma prior, the Horseshoe prior, the Lasso prior and the double gamma prior.}
	\label{fig:incl_theta_EA}
\end{figure}

Figures \ref{fig:incl_beta_EA} and \ref{fig:incl_theta_EA} display the posterior inclusion probabilities for the means of the initial states and the innovation variances of the VAR coefficients, respectively.
A few things about Figure \ref{fig:incl_beta_EA} are striking. First, the posterior inclusion probabilities on the diagonal, meaning those belonging to the parameter of each equation's own autoregressive term, appear to be those that are the highest, while off diagonal elements are more likely to be excluded. Second, the equation for the short-term interest rate is characterized by a large amount of parameters with a high inclusion probability, across all priors. Third, the first lag tends to have higher posterior inclusion probabilities than the second lag, which is in line with the literature. Finally, the triple gamma prior can be seen to often have either the largest or the smallest posterior inclusion probability compared to the other priors, which can be seen as a reflection of the high amount of prior mass placed near
\comment{the shrinkage factors $\rho^\beta_{ij} = 1$ and $\rho^\beta_{ij} = 0$ of $\beta_{ij}^\beta$}, as illustrated in Section \ref{sec_pro_BMA}. This BMA-like behavior yields a prior that is prone to be more absolute when it comes to inclusion decisions.

Now, we shift our focus to the posterior inclusion probabilities for the $\theta^\beta_{ij}$'s plotted in Figure \ref{fig:incl_theta_EA}. Compared to the means of the inital states, almost all inclusion probabilities are essentially zero, with virtually only the triple gamma picking up (faint) signals, in particular with respect to the equations for the financial variables in the model, namely interest rate and nominal exchange rate. This lack of variability is unsurprising, as it is well known (see, e.g., \cite{feldkircher2017sophisticated}) that stochastic volatility in a TVP-VAR model for macroeconomic variables can explain a large part of the variability in the data. Despite this, the triple gamma, thanks to its heavy tails, is still capable of picking up weak signals in the data that the other shrinkage priors we considered are not able to discern from noise.

\begin{figure}[t!]
	\centering
	\includegraphics[width=1\textwidth]{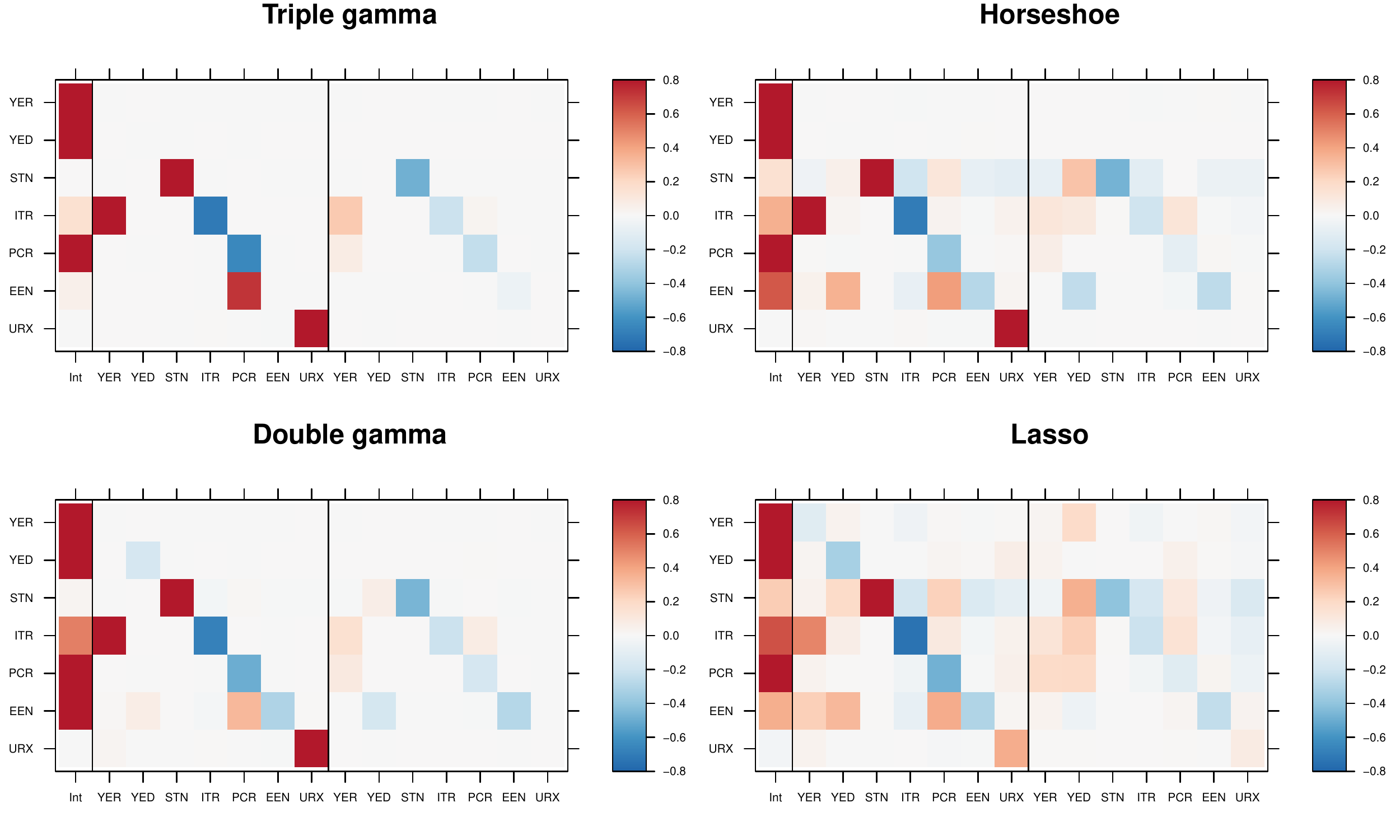}
	\caption{Posterior median of $\beta ^\beta_{ij}$ under the triple gamma, Horseshoe, double gamma and Lasso for the Euro area model. The vertical lines delimit the intercept, first and second lag, respectively.}
	\label{fig:EA_heatmap_beta}
\end{figure}

\begin{figure}[t!]
	\centering
	\includegraphics[width=1\textwidth]{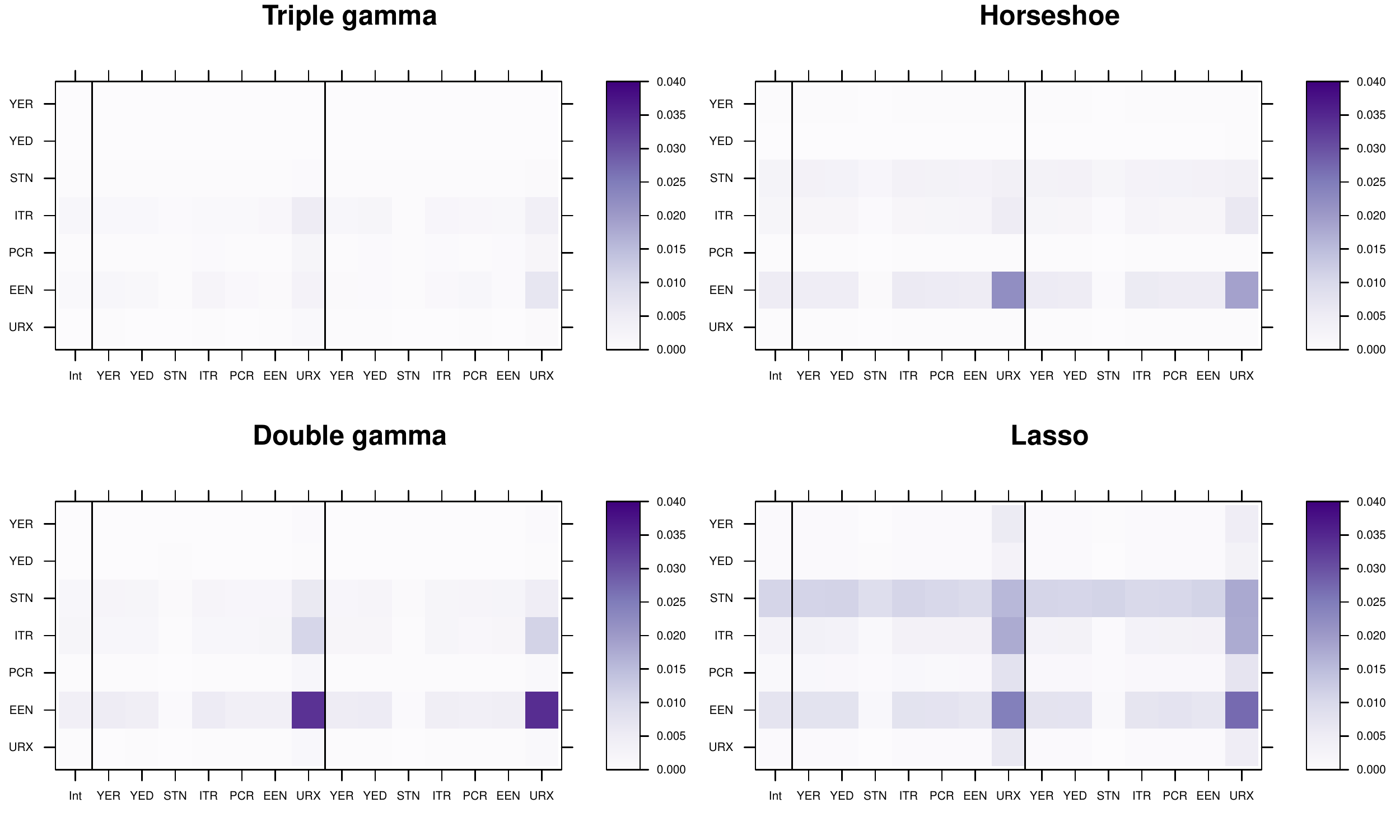}
	\caption{Posterior median of $\left\lvert \sqrt \theta ^\beta_{ij}\right\rvert$ under the triple gamma, Horseshoe, double gamma and Lasso for the Euro area model. The vertical lines delimit the intercept, first and second lag, respectively}
	\label{fig:EA_heatmap_theta}
\end{figure}

Given that the triple gamma tends to include more time variation than the other priors, overfitting might be considered a concern. However, Figures  \comment{\ref{fig:EA_heatmap_beta}  and \ref{fig:EA_heatmap_theta}}  put these fears to rest. They display the posterior median of \comment{$\beta ^\beta_{ij}$ and $\left\lvert \sqrt \theta ^\beta_{ij}\right\rvert$}, respectively. Here the triple gamma can be seen to be quite conservative, both in terms of which parameters to include, as well as their magnitude. In particular the medians of the $\left\lvert \sqrt \theta ^\beta_{ij}\right\rvert$ are interesting, as they are closest to zero under the triple gamma prior, despite having the highest posterior inclusion probabilities among all considered priors, pointing towards the triple gamma's ability to pick up even small signals with a higher degree of confidence than other priors.

In Figures \ref{EA_states1} and \ref{EA_states2} in Appendix \ref{sec:application_app}, all the posterior paths of \comment{$\Phim_{1,t}$ and $\Phim_{2,t}$} under the triple gamma prior are shown.

\section{Conclusion}   \label{conclude}

In the present paper,  shrinkage for time-varying parameter (TVP) models was investigated within a Bayesian framework  with the goal to automatically reduce time-varying  parameters to static ones, if the model is overfitting. This goal was  achieved by suggesting the triple gamma prior as a new shrinkage priors for  the process variances of varying coefficients, extending
previous work  using spike-and-slab priors, 
the Bayesian Lasso, 
or the double gamma prior. 
The triple gamma prior is related to the  normal-gamma-gamma prior applied for variable selection in highly structured regression models \citep{gri-bro:hie}. It contains the well-known Horseshoe prior 
as a special case, however it is more flexible, with two shape parameters that control concentration at zero and the tail behaviour. This leads to a BMA-type behaviour which allows not only variance shrinkage, but also variance  selection.


In our application,  we considered  time-varying parameter VAR models with stochastic volatility.
Overall, our findings suggest that the family of  triple gamma priors introduced in this paper for  sparse TVP models is successful in avoiding overfitting, if  coefficients are, indeed, static or even insignificant.
The   framework  developed in this paper is  very general  and  holds the promise to be useful for introducing  sparsity in  other  TVP   and  state space models  in many different settings. Nevertheless, a number of extensions seem to be worth pursuing.

In particular, in ultra-sparse settings,  modifications seem sensible.
Currently, the hyperprior for the global shrinkage parameter of the triple gamma prior is selected in a way that it implies a uniform prior on \lq\lq model size\rq\rq . A generalization of Theorem~\ref{theo2} would allow to choose hyper priors that induce higher sparsity.
Furthermore, in the variable selection literature, special  priors  such as the Horseshoe+  \citep{bha-etal:hor_est} were suggested for very sparse, ultra-high dimensional settings. Exploiting once more the non-centered parametrization of
a state space model, it is straightforward  to extend this prior to variance selection
using following hierarchical representation:
\begin{align*}
\sqrt{\theta_j} |  \kappa^2 _j, \xiP _j   \sim \Normal{0, \frac{2}{\kappa_B^2}  \kappa^2 _j  \xiP _j },  \quad    \kappa _j  \sim \Studentnu{1},  \quad\xiPsqr _j  \sim \Studentnu{1}.
\end{align*}
We leave both extensions for future research.

An  important limitation of our approach is that shrinking a variance toward zero implies that a coefficient is fixed over the entire observation period of the time-series. In future research we will investigate
dynamic shrinkage priors \citep{kal-gri:tim,kow-etal:dyn,roc-mca:dyn}  
where  coefficients can be both fixed and dynamic.



\vspace{6pt}



\authorcontributions{The authors contributed equally to the work.}


\conflictsofinterest{The authors declare no conflict of interest.}


\appendixtitles{no} 
\appendix

\section{Proofs}  \label{proofs}

\begin{proof}[Proof  of  Theorem~\ref{theo1}.]
To proof Part~(a), rewrite prior (\ref{TRipleII}) in the following way by rescaling   $ \xi_j^2$   and  $\kappa_j^2$:
\begin{align}
\label{TRipleSq}  & \sqrt{\theta_j} |  \xitilde_j ^2 , \kappatilde_j^2, \kappa_B^2 \sim
\Normal{0, \frac{2}{\kappa_B^2} \frac{\xitilde_j ^2}{\kappatilde_j^2}},  \quad
\xitilde_j ^2| a^\xi   \sim \Gammad{a^\xi,  a^\xi}, \quad \kappatilde_j^2| c^\xi  \sim \Gammad{c^\xi,  c^\xi} ,
\end{align}
and use the fact that in (\ref{TRipleSq}) the random variable   $\xiF_j= \xitilde_j^2/\kappatilde_j^2$ follows the F-distribution:
\begin{align*}
\xiF _j   = \frac{\xitilde_j^2}{\kappatilde_j^2} \sim  \frac{\Gammad{a^\xi,  a^\xi}}{\Gammad{c^\xi,  c^\xi}} =_d \Fd{2a^\xi, 2 c^\xi},
\end{align*}
where $p(\xiF_j )$ is given by:
\begin{eqnarray} \label{densF}
p(\xiF_j )=\frac{1}{ \Betafun{a^\xi,c^\xi}}  \!
\left(\frac{a^\xi}{c^\xi} \xiF_j  \right)  ^{a^\xi -1}\! \left( 1+ \frac{a^\xi}{c^\xi} \xiF_j \right) ^{-(a^\xi+c^\xi)}.
\end{eqnarray}
This yields (\ref{repF}).

Using that    $\eta_j= 1/ \xiF _j  \sim \Fd{2 c^\xi, 2a^\xi}$, we obtain from (\ref{repF}) that
\begin{align*}
p(\sqrt{\theta_j} |\kappa^2_B, a^\xi, c^\xi) = & \frac{\sqrt{\kappa_B^2}(c^\xi)  ^{c^\xi}}{\sqrt{4 \pi}( a^\xi) ^{c^\xi} \Betafun{a^\xi,c^\xi}}
\int_0^\infty
\exp\left(-\frac{\theta_j \kappa_B^2  \eta_j}{4}   \right)
\eta_j ^{c^\xi -\frac{1}{2}}  \left(1 +\frac{c^\xi \eta_j}{a^\xi} \right)^{-(a^\xi + c^\xi)} \text{d} \,  \eta_j .
\end{align*}
A  change of variable with  $y_j= c^\xi \eta_j/a^\xi$  proves Part~(b):
\begin{align*}
p(\sqrt{\theta_j} |\phixi, a^\xi, c^\xi) = & \frac{1}{\sqrt{2\pi \phixi} \Betafun{a^\xi,c^\xi}}
\int_0^\infty
\exp \left(-\frac{\theta_j}{2 \phixi}  y_j \right)     y_j ^{c^\xi -\frac{1}{2}}
\left(1 + y_j\right)^{-(a^\xi + c^\xi)} \text{d} \,  y_j \\
= & \frac{\Gamfun{c^\xi+\frac{1}{2}}}{\sqrt{2\pi \phixi} \Betafun{a^\xi,c^\xi}}    \Uhyp{c^\xi + \frac{1}{2}, \frac{3}{2} - a^\xi, \frac{\theta_j}{2 \phixi} },
\end{align*}
where $\phixi=\frac{2 c^\xi}{\kappa_B^2 a^\xi}$.   
\end{proof}

\begin{proof}[Proof  of  Theorem~\ref{theo4}.]

Using \citet[13.5.8]{abr-ste:han}, we obtain  for $a$ and $1 < b <2$ fixed  that $\Uhyp{a,b,z}$ behaves for small $z$ as:
\begin{align*}
\Uhyp{a,b,z}=  \frac{\Gamfun{b-1}}{\Gamfun{a}} z ^{1 - b} + O(1).
\end{align*}
Since $b= 3/2- a^\xi$ in the  expression for  $p(\sqrt{\theta_j} |\phixi, a^\xi, c^\xi)$ given in (\ref{theo1pd}), the condition $1 < b <2$ is equivalent to $0< a^\xi < 0.5$ and this proves Part~(a):
\begin{align*}
p(\sqrt{\theta_j} |\phixi, a^\xi, c^\xi) =   \frac{\Gamfun{\frac{1}{2}- a^\xi }}{\sqrt{\pi}  (2 \phixi) ^{a^\xi} \Betafun{a^\xi,c^\xi}} \left( \frac{1}{\sqrt{\theta_j}} \right)^{1 - 2 a^\xi} + O(1).
\end{align*}
For $b=1$ we obtain from   \citet[13.5.9]{abr-ste:han} that $\Uhyp{a,b,z}$ behaves for small $z$ as follows:
\begin{align*}
\Uhyp{a,b,z}=            - \frac{1}{\Gamfun{a}}  \left( \log z  +  \psi(a) \right)  + O(|z \log z|),
\end{align*}
where $\psi(\cdot)$ is the digamma function.
Since $b=0$ is  equivalent with  $a^\xi = 0.5$, this proves Part~(b):
\begin{align*}
p(\sqrt{\theta_j} |\phixi, a^\xi, c^\xi) =  \frac{1}{\sqrt{2\pi \phixi} \Betafun{a^\xi,c^\xi}}
\left( - \log \theta_j  + \log (2 \phixi) - \psi(c^\xi+{\frac{1}{2}}) \right)
+ O(| \theta_j \log \theta_j|).
\end{align*}
Using  formulas 13.5.10-13.5.12 in \citet{abr-ste:han}, we obtain  for $a$ and $b<1$ fixed  that $\Uhyp{a,b,z}$ behaves for small $z$ as follows:
\begin{align*}
\Uhyp{a,b,z}= \left\{ \begin{array}{ll}
\displaystyle \frac{\Gamfun{1-b}}{\Gamfun{1+a-b}}   + O(z^{1-b}), & 0 < b < 1, \\[1mm]
\displaystyle        \frac{1}{\Gamfun{1+a}}   + O(|z \log z|), &  b =0 ,\\[2mm]
\displaystyle        \frac{\Gamfun{1-b}}{\Gamfun{1+a-b}}  + O(|z|), & b < 0. \\[1mm]
\end{array} \right.
\end{align*}
Since $O(z^{1-b})$ with $ b < 1$, $O(|z \log z|)$ and $O(|z|)$ converge to 0 as $z\rightarrow 0$, we obtain:
\begin{align*}
\lim _{z\rightarrow 0} \Uhyp{a,b,z}  = \frac{\Gamfun{1-b}}{\Gamfun{1+a-b}}.
\end{align*}
This proves Part~(c) as condition $b <1$ is equivalent to $a^\xi > 0.5$:
\begin{align*}
\lim_{\sqrt{\theta_j} \rightarrow 0} p(\sqrt{\theta_j} |\phixi, a^\xi, c^\xi) =
\frac{\Gamfun{c^\xi+{\frac{1}{2}}}}{\sqrt{2\pi \phixi} \Betafun{a^\xi,c^\xi}}  \lim _{z\rightarrow 0} \Uhyp{c^\xi+{\frac{1}{2}}, \frac{3}{2}- a^\xi,z} =
\frac{\Gamfun{c^\xi+{\frac{1}{2}}}\Gamfun{a^\xi - \frac{1}{2}}}{\sqrt{2\pi \phixi} \Betafun{a^\xi,c^\xi}
	\Gamfun{a^\xi + c^\xi }}.
\end{align*}
Finally, using \citet[13.1.8]{abr-ste:han}, we obtain as $z \rightarrow \infty$:
\begin{align*}
\Uhyp{a,b,z}= z^{ -a} \left[1  + O \left( \frac{1}{z}\right) \right], \quad  b < 1.
\end{align*}
Therefore as  $\sqrt{\theta_j} \rightarrow \infty$
\begin{align*}
p(\sqrt{\theta_j} |\phixi, a^\xi, c^\xi) =
\frac{\Gamfun{c^\xi+{\frac{1}{2}}}(2 \phixi) ^{c^\xi} }{\sqrt{\pi} \Betafun{a^\xi,c^\xi}}
\left( \frac{1}{\sqrt{\theta_j}} \right)^{ 2 c^\xi+1} \left[1  + O \left( \frac{1}{\theta_j}\right) \right].
\end{align*}
\end{proof}
\begin{proof}[Proof  of  Lemma~\ref{Lem1}.]
Defining $ \xiFtilde _j=\frac{a^\xi}{c^\xi} \xiF_j$, representation (a) 
follows immediately from   (\ref{repF}) and  (\ref{densF}). 
Representation (b)   is obtained  from (\ref{TRipleSq}) by rescaling   $ \xitilde_j^2$   and  $\kappatilde_j^2$.
To derive representation (c), 
integrate  (\ref{TRipleSq}) 
with  respect to  $\kappatilde_j ^2$, using the common normal-scale mixture representation of the Student-$t$ distribution.
Finally, representation (d)  
is obtain from (\ref{repST}) by rescaling.
\end{proof}

\begin{proof}[Proof  of  Theorem~\ref{theo2}.]
The equivalence of  (\ref{hypfinalproof})  and  (\ref{hypphiproof}) follows immediately from
$ \phixi = (c^\xi/a^\xi) (2/\kappa_B^2) \sim \Betapr{c^\xi,a^\xi } $, since  $2/\kappa_B^2  \sim \Fd{2 c^\xi,2a^\xi }$.
In addition, (\ref{hypphiproof}) implies that
\begin{align*}
\frac{\phixi}{1+ \phixi}   \sim \Betadis{c^\xi,a^\xi }.
\end{align*}
Using  representations (\ref{repBP}) and (\ref{tpbn2}) of the tripe gamma prior,  we can show:
\begin{align*}
\kappashr_j< 0.5
\quad   \Leftrightarrow   \quad  \xi^2_j  >1
\quad   \Leftrightarrow   \quad  \phixi \xiFtilde _j   >1
\quad  \Leftrightarrow    \quad   \frac{1}{1+\xiFtilde _j}   <   \frac{\phixi}{1+ \phixi} ,
\end{align*}
where $\xiFtilde _j\sim \Betapr{a^\xi,c^\xi}$ and, consequently,
\begin{eqnarray*} 
	\frac{\xiFtilde _j}{1+ \xiFtilde _j} \sim \Betadis{a^\xi,c^\xi}  \quad  \Leftrightarrow \quad
	\frac{1}{1+ \xiFtilde _j}  \sim \Betadis{c^\xi,a^\xi }.
\end{eqnarray*}
Hence, $\pi^\xi= \Prob{\kappashr_j< 0.5 }= F_X(Y)$,
where  $F_X$ is the cdf of  a  random variable $X \sim \Betadis{c^\xi,a^\xi }$ and  the random variable $Y \sim \Betadis{c^\xi,a^\xi }$  arises from  the same distribution.
It follows immediately that  $ \pi^\xi  \sim \Uniform{0,1}$.
\end{proof}

\section{Details on the MCMC scheme}  \label{sec_details}

\noindent In Step~(b),
\begin{align*}
p(a^\xi |  \zcond{a^\xi} , \ym)  &\propto
\left(\prod_{j=1}^d p(\sqrt{\theta_j} |\kappaphi_j^2, \phixi)\right)  {p(\kappa_B^2|a^\xi, c^\xi)} p(a^\xi),
\end{align*}
where $p(\kappa_B^2|a^\xi, c^\xi)$ is given by:
\begin{eqnarray*}
	p(\kappa_B^2|  a^\xi, c^\xi)=\frac{1}{2 ^ {a^\xi} \Betafun{a^\xi,c^\xi}}  \!
	\left(\frac{a^\xi}{c^\xi}  \kappa_B^2  \right)  ^{a^\xi -1}\! \left( 1+ \frac{a^\xi}{2 c^\xi} \kappa_B^2 \right) ^{-(a^\xi+c^\xi)}.
\end{eqnarray*}
Therefore,
\begin{align}  \label{postaxi}
p(a^\xi |  \zcond{a^\xi} , \ym) &\propto  \dfrac{2^{- d a^\xi}}{ {\Gamma(a^\xi)}^{d}}  (a^\xi)^{ d (a^\xi + 1/2)/2}
\left(\dfrac{\kappa_B^2 } { c^\xi }\right )^{ d a^\xi/2} \\
\nonumber &  \left ( \prod_{j=1}^d  \kappaphi_j^2 \theta_j \right)^{a^\xi/2}  \left (\prod_{j=1}^d
K_{a^\xi -1/2} \left ( \sqrt{\dfrac{ \kappaphi_j^2 \kappa_B^2 a^\xi } {c^\xi}} |\sqrt{\theta_j} | \right)  \right )  \times\\
\nonumber & 
{\frac{1}{2 ^ {a^\xi} \Betafun{a^\xi,c^\xi}}  \!
	\left(\frac{a^\xi}{c^\xi}  \kappa_B^2  \right)  ^{a^\xi -1}\! \left( 1+ \frac{a^\xi}{2 c^\xi} \kappa_B^2 \right) ^{-(a^\xi+c^\xi)} }
\comment{(2a^\xi)^{\alpha_{a^\xi}-1} (1-2a^\xi) ^ {\beta_{a^\xi} -1} } \\
\end{align}
Hence, $\log q_a ( a^\xi) $ is given by (using $ \Gamfun{a^\xi}= \Gamfun{a^\xi +  1}/a^\xi $):
\begin{align} \label{postaxiq}
\log q_a ( a^\xi) & =  a^\xi \left(-d \log 2 + \dfrac{d}{2} \log \kappa_B^2 - \dfrac{d}{2} \log c^\xi +
\dfrac{1}{2} \sum_{j=1}^d \log \kappaphi_j^2 +  \dfrac{1}{2}\sum_{j=1}^d  \log \theta_j\right)  \\
\nonumber &  + \dfrac{5}{4}d \log a^\xi  + d \dfrac{a^\xi}{2} \log a^\xi  - d \log \Gamma(a^\xi +1) \\
\nonumber & + \sum_{j=1}^d  \log K_{a^\xi -1/2} \left ( \sqrt{\dfrac{ \check\kappa_j^2 \kappa_B^2 a^\xi } {c^\xi}} |\sqrt{\theta_j} | \right)   \qquad \text{(prior on $\theta_j$)} \\
\nonumber &{ - \log \Betafun{a^\xi,c^\xi}
	+ a^\xi   \left(\log a^\xi  + \log ( \frac{\kappa_B^2}{2 c^\xi}) \right)  - \log a^\xi - (a^\xi+c^\xi) \log  \left( 1+ \frac{a^\xi \kappa_B^2}{2 c^\xi}  \right)
	\quad \text{(prior on $\kappa_B^2$)} }\\
&+  (\alpha_{a^\xi}-1) \log(2a^\xi) - (\beta_{a^\xi} -1) \log(1-2a^\xi) \qquad \text{(prior on $a^\xi$)} \\
\nonumber & + \log a^\xi + \log(0.5 - a^\xi)   \qquad \text{(change of variable)}
\end{align}
In Step~(c),
\begin{align} \label{GIGxiproof}
p(\xiphi_j^2 | \zcond{\xiphi_j^2} , \ym)  & \propto p(\sqrt{\theta_j}| \xiphi^2_j, \kappaphi^2_j, \phixi) p(\xiphi^2_j|a^\xi) \\
\nonumber 	 & \propto (\xiphi_j^2)^{-1/2} \exp \left \{  - \dfrac{ \kappaphi_j^2  }{2 \phixi \xiphi_j^2 } \theta_j \right \} \times
(\xiphi_j^2)^{a^\xi -1}   \exp \left \{  - \xiphi_j^2 \right \}\\
\nonumber 		&   = (\xiphi_j^2)^{a^\xi -1/2 -1}  \exp \left \{ - \dfrac{1}{2} \left(\dfrac{\kappaphi_j^2  \theta_j }{\phixi}\dfrac{ 1}{\xiphi_j^2 }  + 2  \xiphi_j^2	\right)\right \},
\end{align}
which is equal to the GIG-distribution given in (\ref{GIGxi}).\footnote{The pdf of the      $\GIG{p,a,b}$-distribution is given by
	\begin{eqnarray*}
		\displaystyle  f(y) = \frac{(a/b)^{p/2}}{2 K_p(\sqrt{ab})} y^{p-1} e^{-\frac{1}{2}(a y + b/y)},
	\end{eqnarray*}
	where $K_p(z)$ is the modified Bessel function.}

\noindent In Step~(d),
\begin{align}   \label{postcxi}
p(c^\xi |   \zcond{c^\xi} , \ym ) & \propto
\left( \prod_{j=1}^d p(\sqrt{\theta_j} |\xiphi_j^2,  c ^\xi ,\kappa_B^2) \right)   {p(\kappa_B^2|a^\xi, c^\xi)} p(c^\xi) \\
\nonumber &  \propto  \left(\prod_{j=1}^d \frac{\Gamma(\frac{2c^\xi + 1}{2})}
{\Gamma(\frac{2c ^\xi}{2})  \left(2c ^\xi \pi\frac{2 \xiphi^2_j}{\kappa_B^2 a ^\xi}\right)^{1/2}}
\left(1 + \frac{1}{2c ^\xi} \frac{\theta_j}{\left(\frac{2 \xiphi^2_j}{\kappa_B^2 a ^\xi}\right)}\right)^{-\frac{2c^\xi + 1}{2}}\right) \times\\
\nonumber  &  {\frac{1}{2 ^ {a^\xi} \Betafun{a^\xi,c^\xi}}  \!
	\left(\frac{a^\xi}{c^\xi}  \kappa_B^2  \right)  ^{a^\xi -1}\! \left( 1+ \frac{a^\xi}{2 c^\xi} \kappa_B^2 \right) ^{-(a^\xi+c^\xi)} }
\comment{(2c^\xi)^{\alpha_{c^\xi}-1} (1-2c^\xi)^{\beta_{c^\xi} -1}}
\end{align}
Hence, $\log q_c ( c^\xi) $ is given by (using $ \Gamfun{c^\xi}= \Gamfun{c^\xi +  1}/c^\xi$):
\begin{align}   \label{logqcxi}
\log q_c ( c^\xi)  &= d \log \Gamma(c^\xi + 0.5)  - d \log \Gamma(c^\xi +  1)
+ \dfrac{d}{2}\log c^\xi  \\
\nonumber &- (c^\xi + 0.5)\left(  \sum_{j=1}^d  \log(4 c^\xi \xiphi_j^2 + \theta_j \kappa_B^2 a^\xi)  - \sum_{j=1}^{d} \log (4 c^\xi \xiphi_j^2) \right)
\qquad \text{(prior on $\theta_j$)}\\
\nonumber &{ - \log \Betafun{a^\xi,c^\xi}
	-  (a^\xi-1)   \log  c^\xi   - (a^\xi+c^\xi) \log  \left( 1+ \frac{a^\xi \kappa_B^2}{2 c^\xi}  \right)
	\quad \text{(prior on $\kappa_B^2$)} }\\
\nonumber &+  (\alpha_{c^\xi}-1) \log(2c^\xi) +  (\beta_{c^\xi} -1)(1-2c^\xi) \qquad \text{(prior on $c^\xi$)} \\
& + \log c^\xi +   \log(0.5 - c^\xi) \qquad \text{(change of variable)}
\end{align}

\noindent In Step~(e),
\begin{align} \label{Gkappaproof}
p(\kappaphi_j^2 | \zcond{\kappaphi_j^2} , \ym ) &  \propto p(\sqrt{\theta_j}| \xiphi^2_j, \kappaphi^2_j, \phixi) p( \kappaphi^2_j|c^\xi)
\\
\nonumber &   \propto (\kappaphi_j^2)^{1/2} \exp \left \{  - \dfrac{ \kappaphi_j^2  }{2 \phixi \xiphi_j^2 } \theta_j \right \} \times
(\kappaphi_j^2)^{c^\xi -1}   \exp \left \{  - \kappaphi_j^2 \right \}\\
\nonumber & = (\kappaphi_j^2)^{1/2 + c^\xi -1}  \exp \left \{ - \kappaphi_j^2  \left( \dfrac{\theta_j }{2 \phixi \xiphi_j^2}  + 1  \right)\right \},
\end{align}
which is   equal to the gamma distribution given in (\ref{Gkappa}).

\noindent In Step~(f), $p(d_2|\zcond{d_2} , \ym  )$  is equal to following gamma distribution:
\begin{align}  \label{d2postproof}
p(d_2|\zcond{d_2} , \ym  )&  \propto p( \kappa_B^2| d_2) p(d_2| a^\xi, c^\xi)\\
\nonumber &  \propto ( d_2)^{a^\xi }  \exp  \left \{  -d_2 \kappa_B^2  \right \} (d_2)^{c^\xi  -1}  \exp  \left \{  - d_2  \dfrac{2 c^\xi}{ a^\xi}  \right \}\\
\nonumber & = (d_2)^{a^\xi + c^\xi -1}  \exp  \left \{  -d_2 \left ( \kappa_B^2  + \dfrac{2 c^\xi}{ a^\xi} \right )\right \},
\end{align}
and 
\begin{align} \label{kappapostproof}
p( \kappa_B^2 |  \zcond{\kappa_B^2} , \ym )&  \propto \prod_{j=1}^d p(\sqrt{\theta_j}| \xiphi^2_j, \kappaphi^2_j, \phixi)  p( \kappa_B^2| d_2)  \\
\nonumber   & \propto (\kappa_B^2)^{d/2} \exp \left \{ -\dfrac{\kappa_B^2 a^\xi}{4 c^\xi}
\sum_{j=1}^d \dfrac{ \kappaphi_j^2  }{\xiphi_j^2 } \theta_j \right \}
\times  { ( \kappa_B^2)^{a^\xi -1}}   \exp \left \{  - d_2  \kappa_B^2 \right \} \\
\nonumber     &  = { (\kappa_B^2)^{d/2 + a^\xi -1}}  \exp \left \{ - \kappa_B^2  \left( \dfrac{a^\xi}{4 c^\xi}  \sum_{j=1}^d\dfrac{ \kappaphi_j^2 }{ \xiphi_j^2 } \theta_j  + d_2  \right)\right \}
\end{align}
which is   equal to the gamma distribution given in (\ref{kappapost}).

For a symmetric triple gamma prior, where $a^\xi= c^\xi$,
Step~(b) is modified in the following way, if Step~(d) is dropped:
\begin{align}  \label{postaxisym}
q_a(a^\xi ) =  p(a^\xi |  \zcond{a^\xi} , \ym)   {\prod_{j=1}^d  p(\kappaphi_j^2 | c^\xi=a^\xi) }
\propto  p(a^\xi |  \zcond{a^\xi} , \ym)  { \frac{1}{\Gamfun{a^\xi}^d} \left (\prod_{j=1}^d  \kappaphi_j^2 \right)^{a^\xi}} ,
\end{align}
where   $p(a^\xi |  \zcond{a^\xi} , \ym)$ is given by (\ref{postaxi}).
If Step~(b) is dropped, then
Step~(d) is modified in the following way:
\begin{align}  \label{postcxisym}
q_c ( c^\xi)  = p(c^\xi |   \zcond{c^\xi} , \ym )   {\prod_{j=1}^d  p(\xiphi_j^2 | a^\xi=c^\xi) }
\propto   p(c^\xi |   \zcond{c^\xi} , \ym )  { \frac{1}{\Gamfun{c^\xi}^d} \left (\prod_{j=1}^d  \xiphi_j^2 \right)^{c^\xi}} ,
\end{align}
where   $p(c^\xi |   \zcond{c^\xi} , \ym ) $ is given by (\ref{postcxi}).

\section{Posterior paths for the simulated data}
\label{sec:simulated_app}
\begin{landscape}
\begin{figure}
	\centering
	\includegraphics{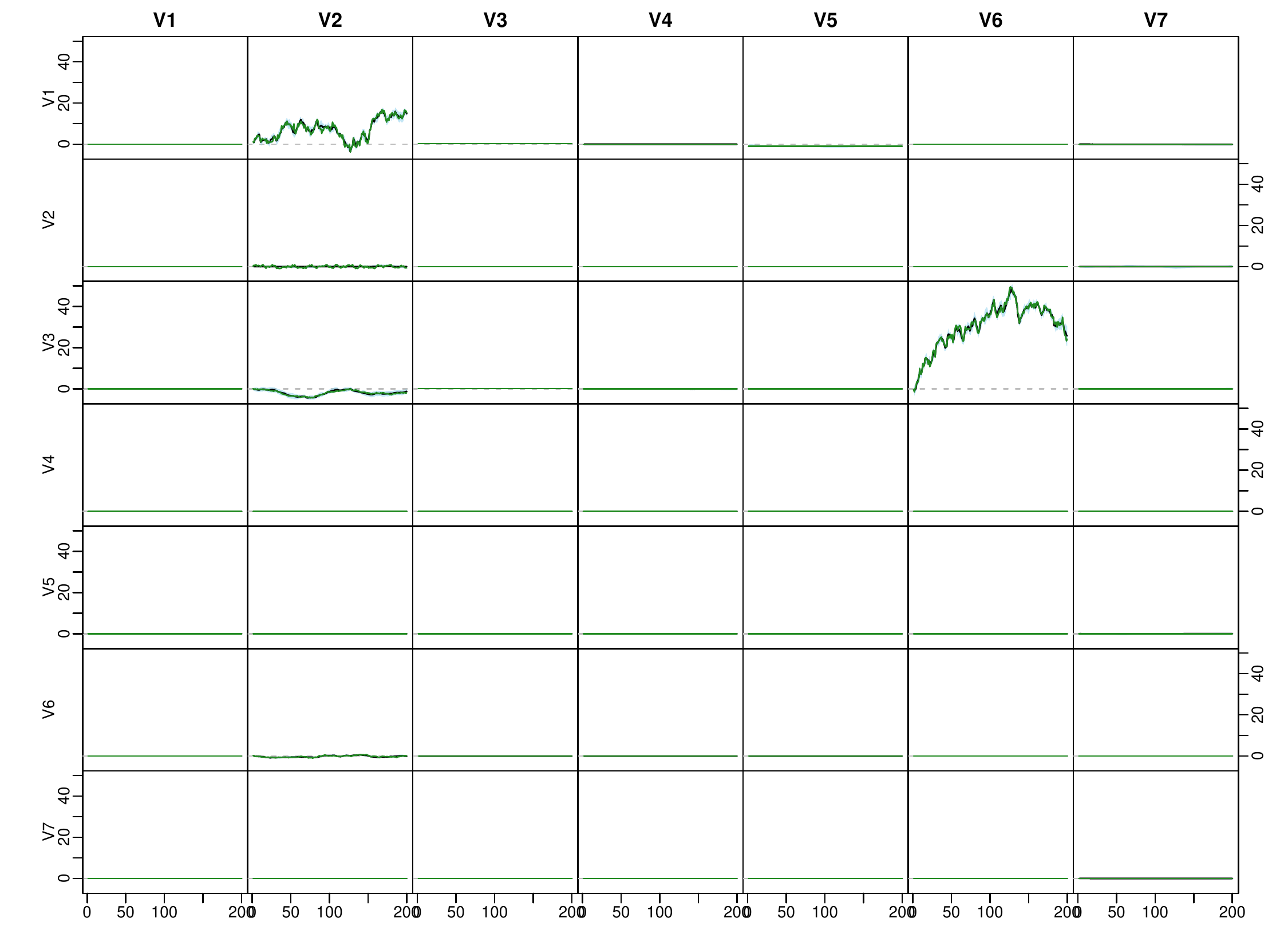}
	\caption{Each cell represents the corresponding state of the matrix \comment{$\Phim_{1,t}$}, for $t = 1, \ldots, T$, for the sparse regime described in Section \ref{sec:example}.
		The solid line is the median and the shaded areas represent $50\%$ and $95\%$ posterior credible intervals.}
	\label{synth_states_sparse}
\end{figure}

\begin{figure}
	\centering
	\includegraphics{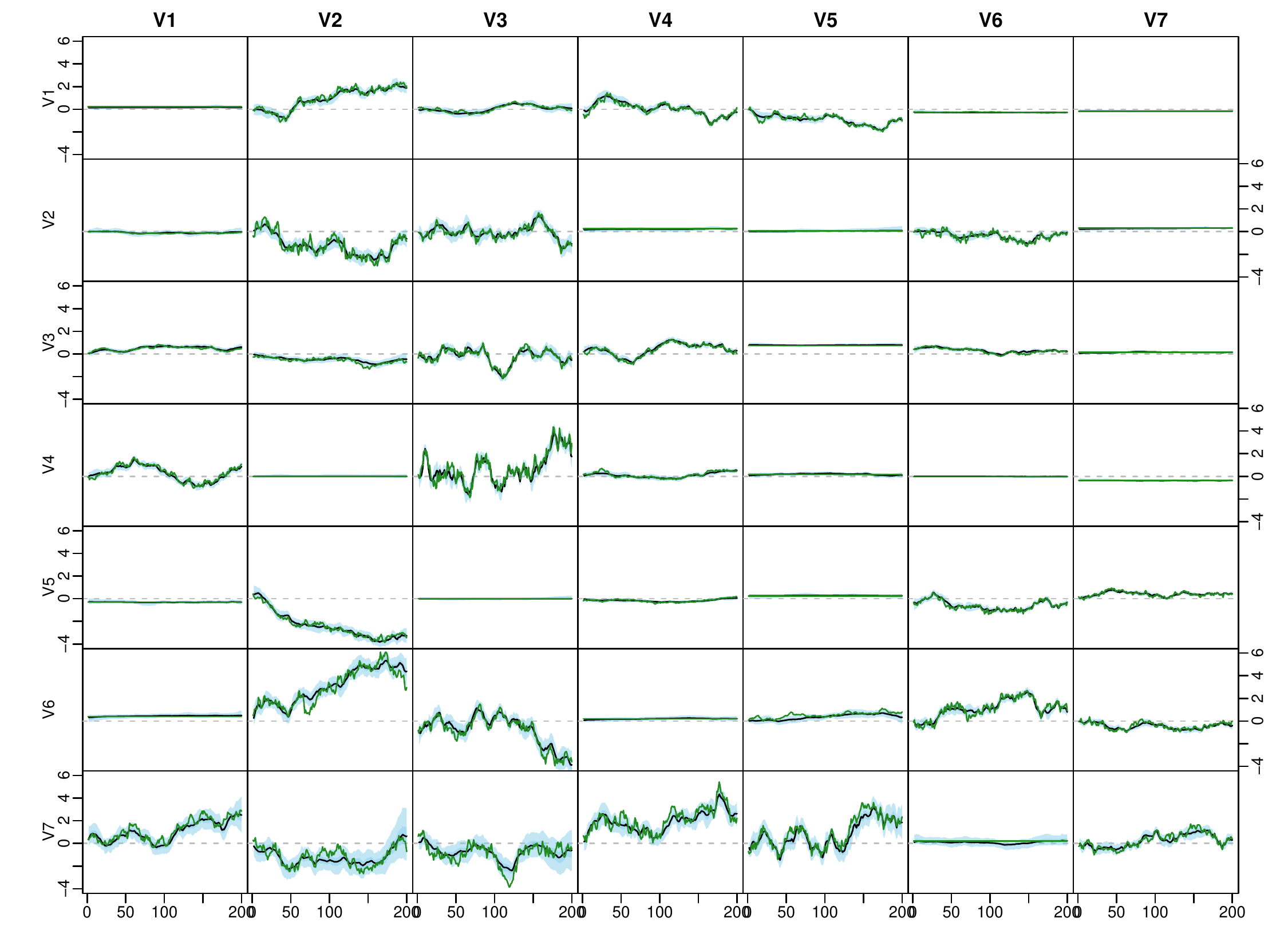}
	\caption{Each cell represents the corresponding state of the matrix $\comment{\Phim_{1,t}}$, for $t = 1, \ldots, T$, for the dense regime described in Section \ref{sec:example}.
		The solid line is the median and the shaded areas represent $50\%$ and $95\%$ posterior credible intervals.}
	\label{synth_states_dense}
\end{figure}

\end{landscape}

\section{Application}
\label{sec:application_app}
\subsection{Data overview}
\begin{table}[H]
	\scriptsize
	\begin{longtable}{l|l|p{9cm}|l}
		\toprule
		Variable                 & Abbreviation & Description                                                                                                                                          & Tcode \\ \midrule
		Real output              & YER          & Gross domestic product (GDP) at market prices in millions of Euros, chain linked volume, calendar and seasonally adjusted data, reference year 1995. & 1     \\
		Prices                   & YED          & GDP deflator, index base year 1995. Defined as the ratio of nominal and real GDP.                                                                    & 1     \\
		Short-term interest rate & STN          & Nominal short-term interest rate, Euribor 3-month, percent per annum                                                                                 & 2     \\
		Investment               & ITR          & Gross fixed capital formation in millions of Euros, chain linked volume, calendar and seasonally adjusted data, reference year 1995.                 & 1     \\
		Consumption              & PCR          & Individual consumption expenditure in millions of Euros, chain linked volume, calendar and seasonally adjusted data, reference year 1995.            & 1     \\
		Exchange rate            & EEN          & Nominal effective exchange rate, Euro area-19 countries vis-à-vis the NEER-38 group of main trading partners , index base Q1 1999.                   & 1     \\
		Unemployment             & URX          & Unemployment rate, percentage of civilian work force, total across age and sex, seasonally adjusted, but not working day adjusted.                          & 2     \\ \bottomrule
	\end{longtable}
	\vspace{0.5em}
	{\raggedright \textbf{Note:} Data was retrieved from \url{https://eabcn.org/page/area-wide-model}. Tcode $= 1$ indicates that differences of logs were taken, while Tcode $= 2$ implies that the raw data was used. \par}
	\caption{Data overview}
	\label{tabdata}
\end{table}

\subsection{Posterior paths}
\begin{landscape}
	\begin{figure}
		\centering
		\includegraphics{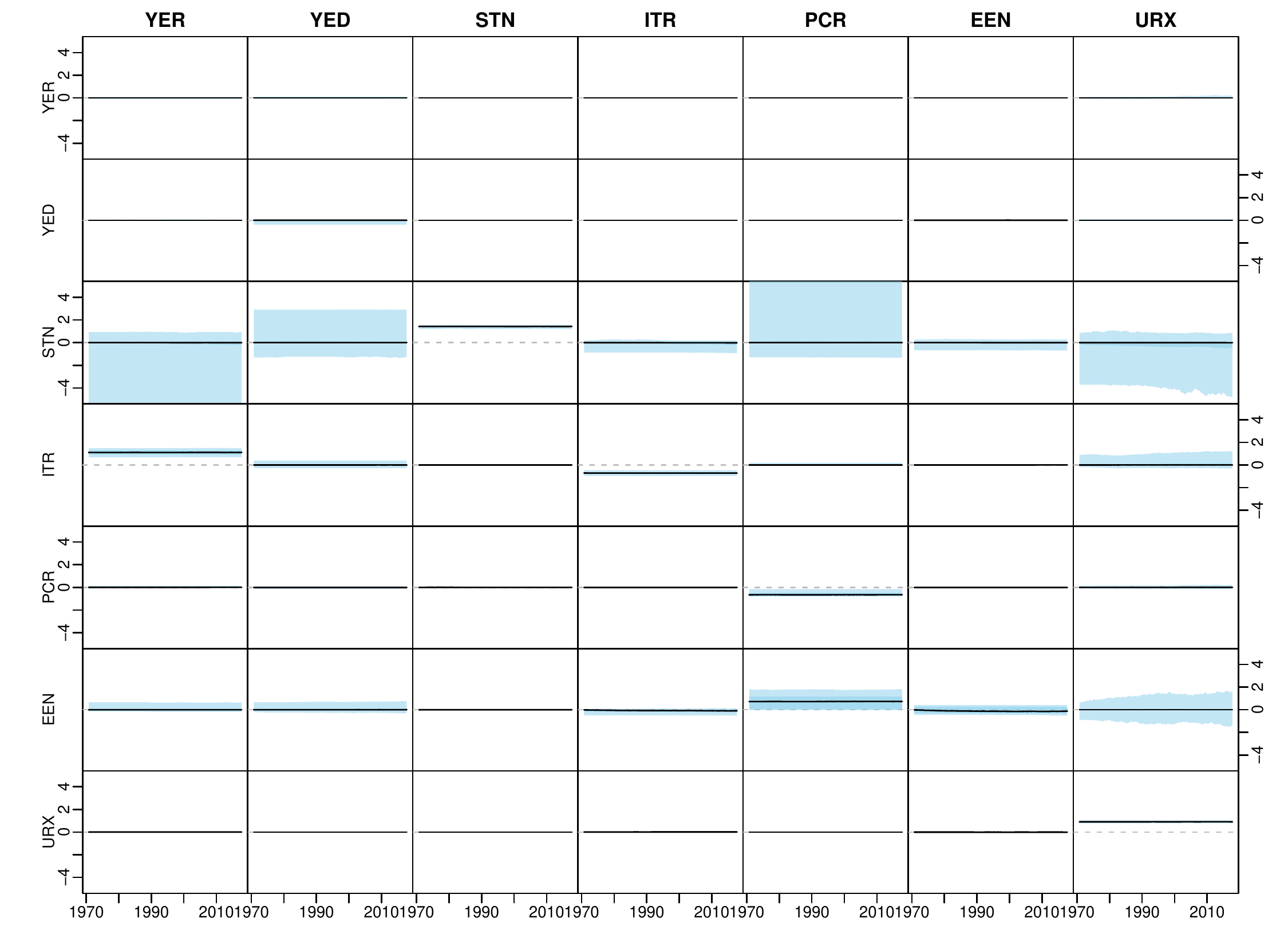}
		\caption{Each cell represents the corresponding state of the matrix $\comment{\Phim_{1,t}}$, for $t = 1, \ldots, T$, for the data described in Section \ref{sec:application}. The solid line is the median and the shaded areas represent $50\%$ and $95\%$ posterior credible intervals.}	
		\label{EA_states1}
	\end{figure}

	\begin{figure}
		\centering
		\includegraphics{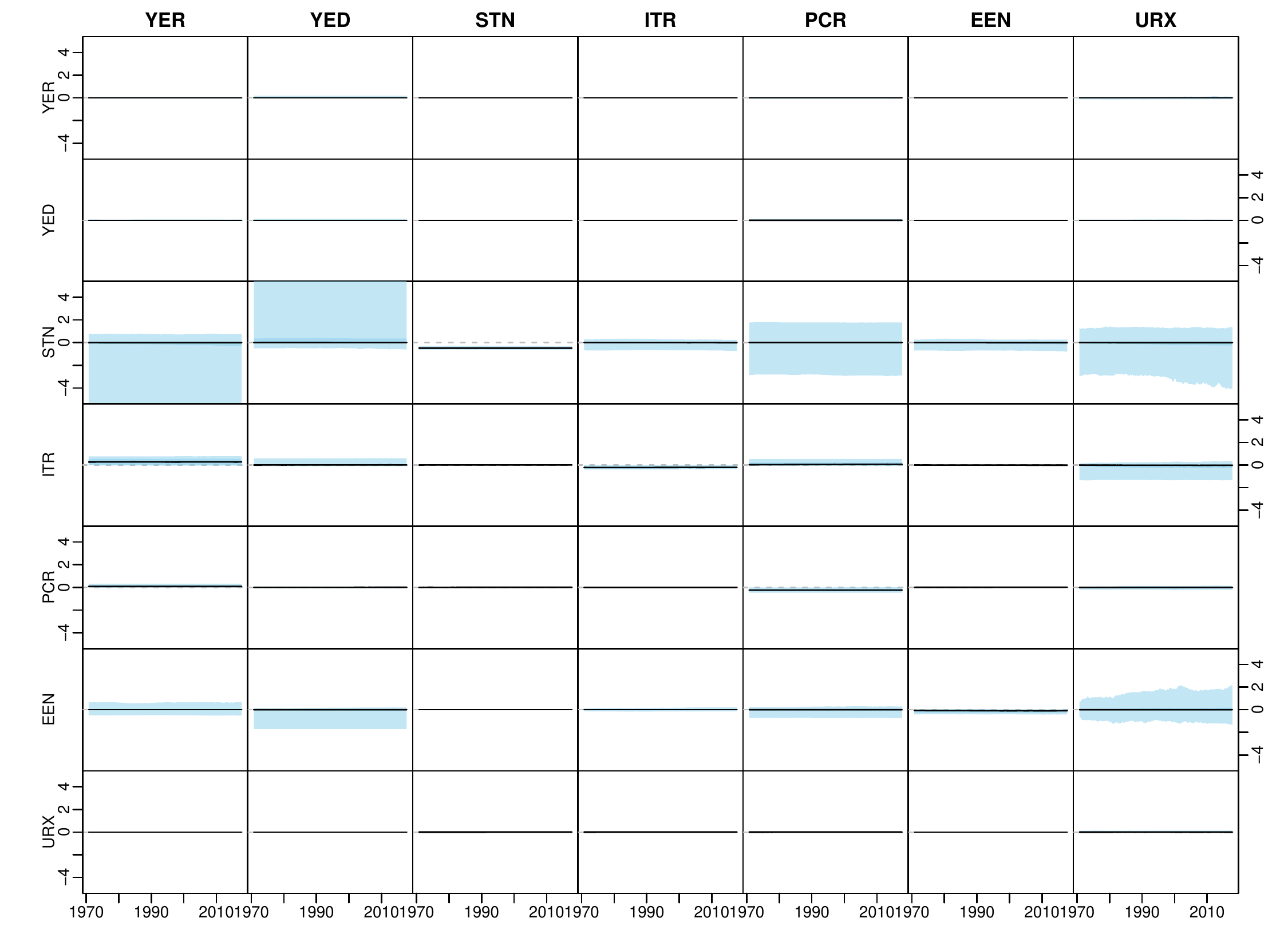}
		\caption{Each cell represents the corresponding state of the matrix $\comment{\Phim_{2,t}}$, for $t = 1, \ldots, T$, for the data described in section \ref{sec:application}. The solid line is the median and the shaded areas represent $50\%$ and $95\%$ posterior credible intervals.}	
		\label{EA_states2}
	\end{figure}
\end{landscape}

\reftitle{References}


\externalbibliography{yes}
\bibliography{sylvia_kyoto_annalisa_peter} 


\sampleavailability{Samples of the compounds ...... are available from the authors.}



\end{document}

%% file: newcommands_triple_gamma.tex
\newcommand{\MVAR}{{\comment{m}}}
\newcommand{\xdum}{x}
\newcommand{\tauij}[2]{\check \tau ^{{#1,2}}_{#2} }
\newcommand{\xiij}[2]{\check \xi ^{{#1,2}}_{#2} }
\newcommand{\kappaij}[2]{\check \kappa ^{{#1,2}}_{#2} }
\newcommand{\lambdaij}[2]{\check \lambda ^{{#1,2}}_{#2} }
\newcommand{\atau}[2]{a ^ {\tau ,{#1}}_{#2}}
\newcommand{\ctau}[2]{c ^ {\tau ,{#1}}_{#2}}
\newcommand{\axi}[2]{a ^ {\xi ,{#1}}_{#2}}
\newcommand{\cxi}[2]{c ^ {\xi ,{#1}}_{#2}}
\newcommand{\kappai}[2]{\kappa ^ {{#1,2}}_{B,#2}}
\newcommand{\lambdai}[2]{\lambda ^ {{#1,2}}_{B,#2}}
\newcommand{\phitaui}[2]{\phi ^ {{\tau,#1}}_{#2}}
\newcommand{\phixii}[2]{\phi ^ {{\xi,#1}}_{#2}}
\newcommand{\e}{\text{e}}
\newcommand{\Uhyp}[1]{U \left(#1\right)}
\newcommand{\kappaphi}{\check \kappa}
\newcommand{\tauphi}{\check \tau}
\newcommand{\lambdaphi}{\check \lambda}
\newcommand{\xiphi}{\check \xi}
\newcommand{\phixi}{\phi^{\xi}}
\newcommand{\phitau}{\phi^{\tau}}
\newcommand{\kappatilde}{\tilde{\kappa}}
\newcommand{\xitilde}{\tilde{\xi}}
\newcommand{\kappashr}{\rho}
 \newcommand{\zcond}[1]{\zm  \minusmcmc{#1}}

 \newcommand{\Phim}{\boldsymbol{\Phi}}

 \newcommand{\btildev}[1]{\tilde{\betav}_{#1}}

\newcommand{\GIG}[1]{\mathcal{GIG} \left(#1\right)}

\newcommand{\xiFsqr}{\psi}
\newcommand{\xiF}{\xiFsqr^2 }
\newcommand{\xiFtilde}{\tilde \xiFsqr^2 }

\newcommand{\xiPsqr}{\xi}
\newcommand{\xiP}{\xiPsqr^2 }

\newcommand{\slab}{\mbox{\tiny slab}}

\newcommand{\Sigmam}{\boldsymbol{\Sigma}} 
\newcommand{\Amul}{\boldsymbol{A}} 
\newcommand{\Dmul}{\boldsymbol{D}} 

\newcommand{\etav}{{\boldsymbol{\eta}}} 

\newcommand{\betac}{\beta} 
\newcommand{\betad}{d} 
\newcommand{\betar}{\boldsymbol{\betac}} 
\newcommand{\betav}{\betar} 
\newcommand{\Xz}{x}
\newcommand{\Xbeta}{{  \mathbf \Xz}}

\newcommand{\zm}{\bm z}

\newcommand{\error}{\varepsilon} 

\newcommand{\errorv}{\boldsymbol{\error}}

\newcommand{\bfz}{{\mathbf{0}}}

\newcommand{\beqnn}{\begin{eqnarray}}
\newcommand{\eeqnn}{\end{eqnarray}}

\newcommand{\kfx}{\betav}  

\newcommand{\kfQ}{{\mathbf{Q}}}  
\newcommand{\kfwc}{ w}
\newcommand{\kfw}{{\mathbf{\kfwc}}}  

\newcommand{\imarg}[2]{^{#1,(#2)}}

\newcommand{\minusmcmc}[1]{_{-#1}} 
\newcommand{\Betafun}[1]{B (#1)}

\newcommand{\Diag}[1]{\mbox{\rm Diag}\left(#1\right)}
\newcommand{\Ew}[1]{\mbox{\rm E}(#1)}   

\newcommand{\Gamfun}[1]{\Gamma (#1)}

\newcommand{\identm}{{\mathbf I}}
\newcommand{\identy}[1]{{\identm}_{#1}}
\newcommand{\indic}[1]{I_{\{#1\}}}
\renewcommand{\indic}[1]{I\{#1\}}

\newcommand{\Probsym}{\mbox{\rm Pr}}
\newcommand{\Prob}[1]{\Probsym (#1)}

\newcommand{\ym}{\bm  y} 

\newcommand{\Betadis}[1]{\mathcal{B}\left(#1\right)}
\newcommand{\Betapr}[1]{\mathcal{BP}\left(#1\right)}

\newcommand{\Bino}[1]{\mbox{\rm BiNom}\left(#1\right)}

\newcommand{\Chisqu}[1]{\chi^2 _{#1}}
\newcommand{\DE}[1]{ \mathcal{DE}\left(#1\right)}
\newcommand{\Dir}[1]{ \mathcal{D}\left(#1\right)}

\newcommand{\Exp}[1]{\mathcal{E}\left(#1\right)}

\newcommand{\Fd}[1]{\mbox{\rm F}\left(#1\right)}

\newcommand{\Gammad}[1]{ \mathcal{G}\left(#1\right)}

 \newcommand{\Normal}[1]{ \mathcal{N}\left(#1\right)}
  \newcommand{\Normaltext}[1]{ \mathcal{N}(#1)}
\newcommand{\Normult}[2]{ \mathcal{N} _{#1}\left(#2\right)}

\newcommand{\Student}[2]{t _{#1} \left(#2\right)}
\newcommand{\Studentnu}[1]{t _{#1}}
\newcommand{\SBeta}[1]{\mbox{SBeta2} \left(#1\right)}
\newcommand{\GG}[1]{\mbox{GG}\left(#1\right)}

\newcommand{\TPB}[1]{\mathcal{TPB}\left(#1\right)}
\newcommand{\Uniform}[1]{\mathcal{U}\left[#1\right]}


%% file: figures_png_papers.tex
{\begin{figure}
\begin{center}
 \scalebox{#4}{\includegraphics{#3.png}}
\caption{#1}\label{#2}
}{\end{center} \end{figure}}

{\begin{figure}
\begin{center}
\scalebox{#5}{\includegraphics{#3.png}}    \scalebox{#5}{\includegraphics{#4.png}}
\caption{#1}\label{#2}
}{\end{center} \end{figure}}

{\begin{figure}
\begin{center}
\scalebox{#6}{\includegraphics{#3.png}}    \scalebox{#6}{\includegraphics{#4.png}} \scalebox{#6}{\includegraphics{#5.png}}
\caption{#1}\label{#2}
}{\end{center} \end{figure}}

{\begin{table}\begin{center}\caption{#1}\label{#2}}{\end{center}\end{table}}